\documentclass[useAMS,usenatbib]{mn2e}
\usepackage{psfig}
\usepackage{multirow}
\usepackage{subfigure}

\title[Weak lensing in modified gravities]{Weak lensing predictions for modified gravities at non-linear scales}
\author[Emma Beynon, David J. Bacon and Kazuya Koyama]{Emma Beynon\thanks{E-mail: emma.beynon@port.ac.uk}, David J. Bacon and Kazuya Koyama\\
Institute of Cosmology and Gravitation, University of Portsmouth, Dennis Sciama Building, Burnaby Road, Portsmouth, PO1 3FX, UK}
\begin{document}

\date{Accepted ---. Received ---; in original form ---}

\pagerange{\pageref{firstpage}--\pageref{lastpage}} \pubyear{2002}

\maketitle

\label{firstpage}

\begin{abstract}
We present a set of predictions for weak lensing correlation functions in the context of modified gravity models, including a prescription for the impact of the nonlinear power spectrum regime in these models. We consider the DGP and $f(R)$ models, together with dark energy models with the same expansion history. We use the requirement that gravity is close to GR on small scales to estimate the non-linear power for these models. We then calculate weak lensing statistics, showing their behaviour as a function of scale and redshift, and present predictions for measurement accuracy with future lensing surveys, taking into account cosmic variance and galaxy shape noise. We demonstrate the improved discriminatory power of weak lensing for testing modified gravities once the nonlinear power spectrum contribution has been included. We also examine the ability of future lensing surveys to constrain a parameterisation of the non-linear power spectrum, including sensitivity to the growth factor $\gamma$.
\end{abstract}

\begin{keywords}
Gravitation; Gravitational Lensing; Cosmology: Theory
\end{keywords}

\section{Introduction}\label{intro}

Consistent observational evidence from various cosmological probes shows that the Universe is currently undergoing a period of accelerated expansion. The observed expansion history can be explained using some form of dark energy or a cosmological constant; however this cosmological constant cannot be explained with current particle physics due to its very small value. An alternative approach is to invoke a modification of gravity; there are many different ways that gravity and/or the equation of state of the dark energy can be modified to allow for the expansion history observed. This makes it impossible to differentiate between the effects of modified gravity and dark energy by measuring the background expansion history alone. However, modifying gravity also produces a distinct growth rate of structure; thus the expansion history and growth history together can be used to distinguish between various models of gravity. This consistency relation to test GR has been proposed and explored by many papers \citep{Uzan:2000mz,Lue:2003ky,Lue:2004rj,Ishak:2005zs,Kunz:2006ca,Chiba:2007rb,Wang:2007fsa,Bertschinger:2008zb,Jain:2007yk,Daniel:2008et,Song:2008vm}. Upcoming weak lensing surveys such as DES\footnote[1]{http://www.darkenergysurvey.org}, Pan-STARRS\footnote[2]{http://pan-starrs.ifa.hawaii.edu} and LSST\footnote[3]{http://www.lsst.org}, and future space surveys such as Euclid\footnote[4]{http://www.ias.u-psud.fr/imEuclid}, will allow a combination of growth of structure and expansion history to be probed to considerably higher precision, which will allow many gravity models to be excluded.

There has been a great deal of work showing how to use weak lensing to discriminate between different gravity models; however this has been restricted to probing the linear regime of the matter power spectrum \citep{Afshordi:2008rd,Schmidt:2008hc,Song:2008xd,Thomas:2008tp,Tsujikawa:2008in,Zhao:2009fn,Zhao:2008bn} or uses methods that do not obtain GR at small scales \citep{Knox:2006fh,Heavens:2007ka,Yamamoto:2007gd,Amendola:2007rr}. The non-linear regime provides much of the power for lensing and can be most easily probed by current and upcoming lensing surveys. This paper examines the effect of including the non-linear regime in modified gravity lensing predictions, including the small-scale GR limit, to see how useful weak lensing will be overall when trying to determine the correct model of gravity. First we look at DGP and $f(R)$ gravity models as examples, and investigate weak lensing's ability to differentiate between these models and dark energy models. We then take a more phenomenological point of view, by parameterising the shape of the matter power spectrum and examining the sensitivity of weak lensing observables to changes to the matter distribution when the expansion history is the same for each model considered. Using these parameters we show how strongly a ground-based survey similar to DES and a space-based survey such as Euclid will be able to discriminate between different growth histories with identical expansion histories.

This paper is organised as follows. In section \ref{lensmodgrav} we briefly describe the DGP and $f(R)$ models of gravity and how they compare with dark energy models. We describe how we calculate matter power spectra for these models, including the GR small-scale limit. We also describe how we proceed to calculate weak lensing observables from these power spectra. In section \ref{results} we present the resulting lensing correlation functions, including realistic errors for future surveys taking into account shape measurement noise and cosmic covariance.
In section \ref{parameterisation} we take the alternative approach of parameterising the non-linear power spectrum, and we investigate how sensitive weak lensing is to these parameters which go beyond the usual growth parameter. We present our conclusions in section \ref{conclusions}.

Throughout this paper we will use a flat cosmology with the WMAP5+SNe+BAO best fit cosmological parameters, which are determined by the background evolution of the Universe. We use a $\Lambda$CDM background for both $\Lambda$CDM and $f(R)$, in which case we take $n_{\rm s}=0.96$, $h=0.71$, $\Omega _{\rm m}=0.27 \pm 0.02$ and $\sigma_8=0.81 \pm 0.03$ \citep{Komatsu:2008hk}. When we use a DGP background, we have $n_{\rm s}=0.998$, $h=0.66$, $\Omega _{\rm m}=0.26 \pm 0.02$ \citep{Fang:2008kc} giving a $\sigma_8=0.66 \pm 0.03$ for an equivalent $\Lambda$CDM model.

\section{Lensing in DGP and $\lowercase{f}(R)$ models}\label{lensmodgrav}

\subsection{Modified gravity power spectra}\label{modified gravity}

As we have already mentioned, there are two key phenomena to model in any gravity in order to calculate the matter power spectrum: the expansion history, quantified by the evolution of the Hubble parameter, and the growth history, quantified by the evolution of density perturbations $\delta$ in the Universe.

For $\Lambda$CDM, the expansion history is given by the Friedmann equation

\begin{equation}
	H^2=\frac{\Omega _{\rm m}}{a^3}+\Omega _\Lambda,
\end{equation}
where $H = \frac{da}{dt} /aH_0$, $a$ is the scale factor and $H_0$ is the present day Hubble constant.

The growth history is described by the density perturbation evolution equation together with the Friedmann equation. At this point we will limit ourselves to the regime where density perturbations evolve linearly.
In this regime we have
\begin{equation}
	\delta '' + \left(\frac{3}{a} + \frac{H '}{H}\right) \delta ' = \frac{3\tilde{G}_{\rm eff}}{2 H^2 a^2} \frac{\Omega _{\rm m}}{a^3} \delta,
\end{equation}
where primes denote differentiation with respect to $a$.  This equation is valid for both dark energy and modified gravity models, where $\tilde{G}_{\rm eff}$ is the effective gravitational constant normalised by the gravitational constant $G$; hence $\tilde{G}_{\rm eff}=1$ for dark energy models, while for modified gravity models

\begin{equation}
\tilde{G}_{\rm eff}=1+\frac{1}{3\beta},
\end{equation}
where $\beta$ is determined by the model.

In this paper, we consider DGP \citep{2000PhLB..485..208D} and $f(R)$ as examples of modified gravity models, as the non-linear power spectra have been studied in great detail in these two models using perturbation theory and N-body simulations. For some reviews of modified gravity models see \cite{Nojiri:2006ri,Durrer:2007re,Koyama:2007rx}.

In DGP, spacetime has five dimensions, while we live on a 4D brane in the 5D bulk. Standard Model particles are bound on the 4D brane, as is gravity on small scales; however on large scales gravity leaks off the brane causing late time acceleration. The scale of the transition from 4D to 5D gravity is governed by the crossover scale, $r_{\rm c}=(1-\Omega _{\rm m})^{-1}$. The extra dimension contributes a further term to the Friedmann equation whose amplitude is governed by $r_{\rm c}$:
\begin{equation}
	H^2-\frac{H}{r_{\rm c}}=\frac{\Omega _{\rm m}}{a^3}.
\end{equation}
The growth history is also altered, giving \citep{Koyama:2005kd}
\begin{equation}
 \beta = 1 - 2Hr_{\rm c} \left(1 + \frac{a H'}{3H} \right).
\end{equation}
In $f(R)$ gravity models the Einstein-Hilbert action is modified to include an arbitrary function of the Ricci scalar, $R$.
In this study we use an $f(R)$ function of the form \citep{Hu:2007nk}
\begin{equation}
	f=-6\Omega _\Lambda - \frac{R_{\rm 0}^2}{R} f_{\rm R_0},
\end{equation}
where $R$ is the Ricci scalar, $R_0$ is the present day Ricci scalar and $f_{\rm R_0}=\left.\frac{df}{dR}\right|_{ R=R_0}$. We use $\left|f_{\rm R_0}\right| = 10^{-4}$, which has been found to fit with cluster constraints \citep{Schmidt:2009am}, to give a background evolution which is approximately $\Lambda$CDM to sub-percent level. This allows us to use the $\Lambda$CDM Friedmann equation and only alter the density evolution equation \citep{Lue:2004rj, Zhang:2005vt, Koyama:2009me} with
\begin{equation}
\beta=1+\frac{1}{3c^2\frac{d^2f}{dR^2}}\left(\frac{a}{\bar{k}}\right)^2,
\end{equation}
where $\bar{k}$ is the dimensionless wavenumber defined as $k(c/H_0)$, $k$ is the wavenumber and $c$ is the speed of light.

For modified gravity to agree with solar system observations it must approach a GR solution on small scales. This means that the non-linear power spectrum must be an interpolation of the modified gravity non-linear power spectrum with no mechanism to obtain the GR result on small scales, $P_{\rm non-GR}(k,z)$, and the GR non-linear power spectrum with the same expansion history as the modified gravity model, $P_{\rm GR}(k,z)$. A fitting formula for this interpolation was proposed by \cite{Hu:2007pj}:
\begin{equation}\label{eq:PPF}
	P(k,z) = \frac{P_{\rm non-GR}(k,z)+c_{\rm nl}(z)\Sigma ^2(k,z) P_{\rm GR}(k,z)}{1+c_{\rm nl}(z)\Sigma ^2(k,z)},
	\label{eq:husaw}
\end{equation}
where $\Sigma ^2(k,z)$ picks out non-linear scales and $c_{\rm nl}(z)$ determines the scale at which the power spectrum approaches the GR result as a function of redshift.

In this paper, we use the fitting fomulae for $\Sigma ^2(k,z)$ and $c_{\rm nl}(z)$ obtained by perturbation theory \citep{Koyama:2009me} and confirmed by N-body simulations \citep{Oyaizu:2008tb,Schmidt:2009sg},
\begin{equation}
\Sigma ^2(k,z)=\left(\frac{k^3}{2\pi^2}P_{\rm lin}(k,z)\right)^{\alpha_1},
\quad c_{\rm nl}(z)=A(1+z)^{\alpha_2}.
\label{sigma}
\end{equation}
where
$P_{\rm lin}(k,z)$ is the modified gravity linear power spectrum.
The non-linear power spectrum for both the $P_{\rm non-GR}$ and $P_{\rm GR}$ is found using the \cite{Smith:2002dz} fitting formula from the linear power spectrum.
For DGP, $A=0.3$, $\alpha_1=1$ and $\alpha_2=0.16$ and for $f(R)$ with $f_{\rm R_0}=10^{-4}$ we use $A=0.08$, $\alpha_1=1/3$ and $\alpha_2=1.05$ for $0\leq z \leq 1$. 
It should be noted these values are not valid for all $\Omega_{\rm m}$ and $\sigma_8$. However, in DGP, these values depend on $\Omega_{\rm m}$ and $\sigma_8$ very weakly, so within our priors for $\Omega_{\rm m}$ and $\sigma_8$ we can assume the values are constant.

We should also emphasise that these fits are confirmed only up to $k=1h$/Mpc due to the lack of resolution in N-body simulations, so we are extrapolating the fits beyond this regime. Clearly it is necessary to check the validity of this extrapolation using N-body simulations with higher resolution 
(see \cite{Schmidt:2008tn} for a different approach using the halo model). However, since the modified gravity power spectrum should approach the GR non-linear power spectrum with the same expansion history, and since the fitting formula (\ref{eq:husaw}) ensures this, our extrapolation is justified.

In applying this formalism, we found that although $f(R)$ fits the N-body results at small $k$, it failed to converge with $\Lambda$CDM at larger $k$ if $\alpha_1=1/3$. This is due to the strong scale dependence of the linear power spectrum, such that $P_{\rm non-GR}$ deviates from $P_{\rm GR}$ strongly on small scales and equation (\ref{eq:PPF}) with $\alpha_1= 1/3$ fails to converge with $P_{\rm GR}$. Thus, we also consider $\alpha_1=1$ and $\alpha_1=2$ cases for $f(R)$ which have more physical behaviour at high $k$.

Since we are interested in how sensitive weak lensing is to different growth histories with the same expansion history, we will also consider a quintessence cold dark matter (QCDM) model. In this case, the equation of state of the dark energy is altered to match the expansion history of DGP, while the density perturbation evolution equations are the same as $\Lambda$CDM.

\begin{figure}
\psfig{figure=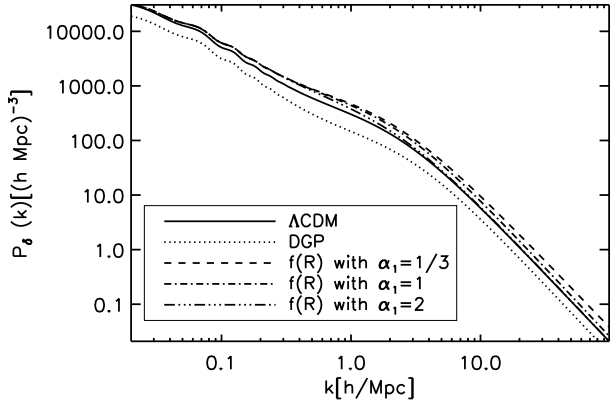,width=84mm}
 \caption{Matter power spectrum for $\Lambda$CDM, DGP and $f(R)$ at z=0.}
 \label{fig:p_delta}
\end{figure}

\begin{figure}
\psfig{figure=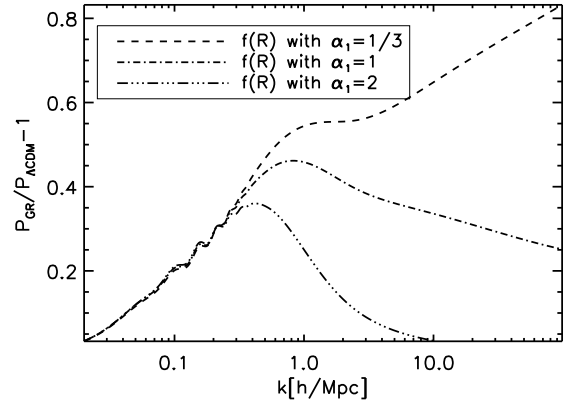,width=84mm}
 \caption{Relative difference between matter power spectra for $\Lambda$CDM and $f(R)$ at $z=0$ for different $\alpha_1$.}
 \label{fig:p_delta diff}
\end{figure}

We show examples of the resulting matter power spectra in Figure \ref{fig:p_delta}.
$f(R)$ models show the scale dependent enhancement of the power spectrum in the linear regime compared with $\Lambda$CDM. For $\alpha_1=1/3$, which fits N-body results well up to $k=1h$/Mpc, the power spectrum fails to converge with $\Lambda$CDM. On the other hand, the power spectrum with $\alpha_1=2$ shows clear convergence; this is shown more explicitly in Figure \ref{fig:p_delta diff}.

\begin{figure}
\psfig{figure=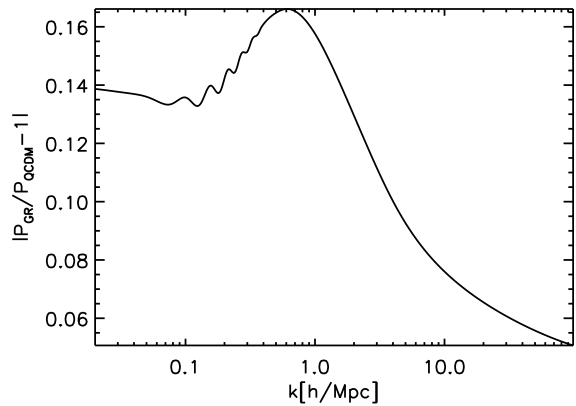,width=84mm}
 \caption{Relative difference between the matter power spectra for DGP and the QCDM model at $z=0$.}
 \label{fig:p_delta_qcdm}
\end{figure}

We also show a comparison between DGP and QCDM power in Figure \ref{fig:p_delta_qcdm}, including our non-linear prescription. In the linear regime the DGP power spectrum receives scale independent suppressions, but it converges to the QCDM power spectrum on non-linear scales due to our inclusion of the GR asymptote.

\subsection{Weak Lensing}\label{weak lensing}

The mass distribution described by the matter power spectrum deflects light all the way along the path from source to observer, so any changes in the matter power spectrum alter the observed image distortion of galaxies. Light bundles are transformed by a shear with two components, $\gamma=\gamma_1+i\gamma_2$, and an isotropic dilation, the convergence $\kappa$ \citep[e.g.][]{Bartelmann:1999yn}. The Jacobian mapping from the unlensed image to the distorted image is given by
\begin{equation}
	A=
	\left(\begin{array}{cc}
	1-\kappa-\gamma_1 & -\gamma_2 \\
	-\gamma_2 & 1-\kappa+\gamma_1
	\end{array}\right).
\end{equation}
Since each galaxy has some unknown intrinsic shape, the amount of lensing cannot be estimated by a single source; however one can correlate the shear estimators of many sources, in which case randomly oriented intrinsic ellipticities will average out, leaving the gravitational shear signal. This will not succeed if galaxy ellipticities are physically aligned, which they are to some degree \citep[e.g.][]{2009ApJ...694..214O}; however, for our purposes we will assume that the resulting physical correlation signal can be removed, leaving only the lensing signal. Here we will concentrate on the convergence correlation function, equal to the sum of the shear correlation functions, which is related to the convergence power spectrum by 
\begin{equation}
	C_\kappa(\theta) = \int ^\infty _0 dl\frac{l}{2\pi}P_\kappa (l)J_0(l\theta)
	\label{eq:ckappa},
\end{equation}
where $\theta$ is the angular distance between the correlated sources and $l$ is the angular wavenumber. The convergence power spectrum is related to the matter power spectrum by \citep[e.g.][]{Bartelmann:1999yn}
\begin{equation}
	P_\kappa (l) = \frac{9}{4} \left(\frac{H_0}{c}\right)^4 \int ^{\chi _{\rm H}} _0 d \chi W(\chi)^2 \frac{P_\delta \left(\frac{l}{\chi},\chi\right)}{a^2},
\end{equation}
with
\begin{equation}
	W(\chi)=\int^{\chi _{\rm H}} _\chi d\chi' G(\chi')\left(1-\frac{\chi}{\chi'}\right),
	\label{eq:w}
\end{equation}
where $\chi$ is comoving distance, $\chi _{\rm H}$ is the comoving distance to the horizon and $G(\chi)$ is the normalised distribution of the sources in comoving distance, corresponding to a redshift distribution. Equation (\ref{eq:w}) is valid for flat cosmologies, which are all that are considered in this paper. 

\begin{figure}
\psfig{figure=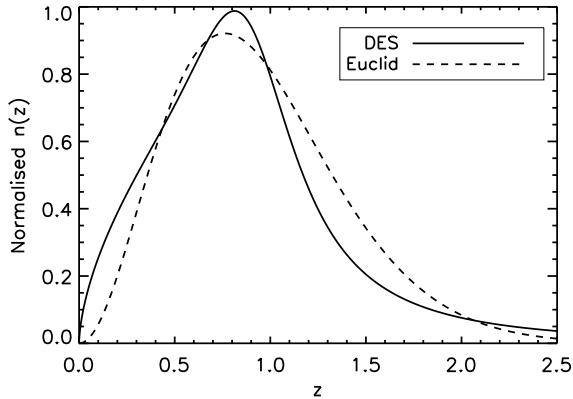,width=84mm}
 \caption{Redshift distributions used for survey predictions: for ground-based survey with $z_m=0.825$, and for Euclid with $z_m=0.91$.}
 \label{fig:redshiftDistribution}
\end{figure}

We will calculate results for realistic notional surveys: a ground-based survey similar to that of the Dark Energy Survey (DES), and a space-based survey such as that of Euclid, using redshift distributions shown in Figure \ref{fig:redshiftDistribution}; the redshift distribution for our ground-based survey was chosen to be the same as for CFHTLS given by \cite{Fu:2007qq} giving a median redshift, $z_m$, of 0.825 and for Euclid we used the distribution given by \cite{Hawken:2009fr} giving $z_m=0.9$.

This can be extended to cross-correlate sources at different redshifts \citep[e.g.][]{Bacon:2004ht,Massey:2007gh} to obtain
\begin{equation}
	P_\kappa (l) = \frac{9}{4} \left(\frac{H_0}{c}\right)^4 \int ^{\chi _{\rm H}} _0 d \chi W_1(\chi) W_2(\chi) \frac{P_\delta \left(\frac{l}{\chi},\chi\right)}{a^2},
\end{equation}
where $W_i$ include the galaxy distributions $G_i$ appropriate for the $i$th redshift bin. This equation together with equation (\ref{eq:ckappa}) relates the matter power spectra from our gravity models to the predicted lensing signal; we will now use these tools to calculate lensing predictions for our models.

It should be noted that in our analysis we have not included the effects of baryons, which are known to have an effect on the matter power spectrum for $k \geq 1 h Mpc^{-1}$ as shown in \cite{White:2004kv,Zhan:2004wq,Jing:2005gm,Rudd:2007zx,Hearin:2009hz}. For non-radiative gas simulations this changes the amplitude of the matter power spectrum by a few percent, however if gas cooling and star formation are included this effect could be considerably larger. Therefore for a full lensing analysis in the non-linear regime these effects must be included as well, but since we must recover GR on small scales, results from gas simulations for GR can be used to refine the models we present.

\section{Results}\label{results}

We calculate the convergence (combined shear) correlation function of equation (\ref{eq:ckappa}) for all of our models, and estimate measurement errors due to intrinsic ellipticity, for the notional ground-based and Euclid surveys using bins with error $\sigma_{\rm shape} = \sqrt{2} \sigma^2_\gamma/\sqrt{N_{\rm pairs}(\theta,\Delta \theta)}$, where $\sigma_\gamma=0.3$. The errors were estimated using $13.3 \textrm{ galaxies arc min}^{-2}$ and a survey area of 5000 square degrees for our ground-based survey (as is appropriate for DES), while for Euclid we use $35 \textrm{ galaxies arc min}^{-2}$ and 20000 square degrees. The covariance matrix for the intrinsic ellipticity noise is diagonal for bins in redshift and angular separation \citep[c.f.][]{2003MNRAS.344..673B}.

 We also include the covariance due to sample variance due to the cosmic matter distribution, $C_{\rm cos}$, which is estimated using the Horizon simulation \citep{Teyssier:2009zd}. 3-D convergence maps were calculated from the 3-D overdensity field, for 75 patches of area 2 square degrees; convergence correlation functions were then measured in each patch. The covariance between the resulting patch correlation functions was measured as an estimator of the true covariance, in 8 angular separation bins logarithmically spaced from $1'$ to $90'$ and in 3 redshift bins (leading to 6 redshift pair bins). 
The diagonal elements of the covariance matrix for mean correlation functions are measured to be approximately $10^{-11}-10^{-9}$ per square degree, making the sample covariance the dominant source of error for larger angles and higher redshifts for both our ground-based survey, with diagonal element values of $10^{-15}-10^{-13}$ and Euclid, $10^{-15}-10^{-13}$. These should be compared with shape noise covariance contributions of $10^{-15}-10^{-11}$ for ground-based and $10^{-16}-10^{-12}$ for Euclid. The covariances were included in our $\chi^2$ estimations using \citep[c.f.][]{Hartlap:2006kj}
\begin{equation}
	\chi ^2=\sum_{i,j}{(d_i-t_i)\left(\frac{n_{\rm o}-1}{n_{\rm o}-n_{\rm b}-2}C_{\rm cos}+\sigma_{\rm shape}^2\right)^{-1}_{ij} (d_j-t_j)},
\end{equation}
where $d$ is the `data', here the fiducial $\Lambda$CDM correlation function in redshift and angular separation bins; $t$ is the alternative gravity model correlation function in those bins, $n_{\rm o}=75$ is the number of realisations of correlation functions used in the calculation of $C_{\rm cos}$ and $n_{\rm b}=48$ is the total number of bins in angular separation and redshift. Note that we use the sample covariance estimate from Horizon (which follows $\Lambda$CDM) for both $\Lambda$CDM and QCDM cases; the QCDM error bars should therefore only be considered as the correct order of magnitude.

We calculate for each of our models the difference in $\chi^2$ between the modified gravity model and a dark energy model (either $\Lambda$CDM or QCDM), applying WMAP+SNe+BAO priors. Note that for $\Lambda$CDM and $f(R)$, we used the $\Lambda$CDM background \citep{Komatsu:2008hk} and for DGP and QCDM, the DGP background \citep{Fang:2008kc} was used (see \S1). 

\begin{figure*}
\centering
\mbox{
\subfigure[Including non-linear effects for sources with $z_{\rm m}=0.825$, with ground-based survey errors]{\psfig{figure=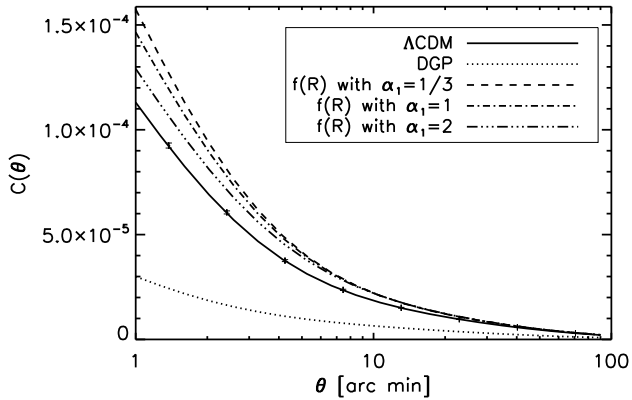,width=84mm}} \quad
\subfigure[Including non-linear effects for sources with $z_{\rm m}=0.9$ with Euclid errors]{\psfig{figure=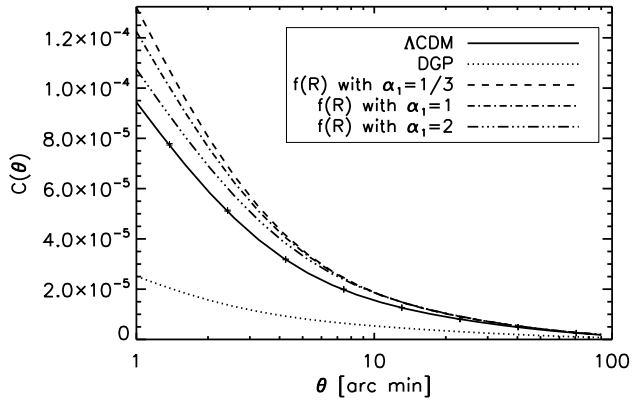,width=84mm}}
}
\mbox{
\subfigure[Not including non-linear effects for sources with $z_{\rm m}=0.825$ with ground-based survey errors]{\psfig{figure=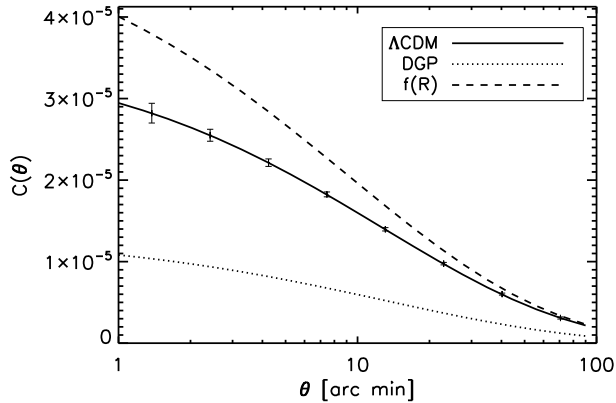,width=84mm}} \quad
\subfigure[Not including non-linear effects for sources with $z_{\rm m}=0.9$ with Euclid errors]{\psfig{figure=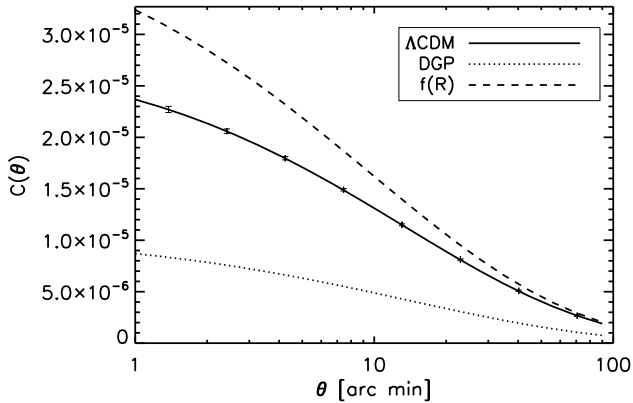,width=84mm}}
}
 \caption{Correlation function predicted for $\Lambda$CDM, DGP and $f(R)$ with error estimates for ground-based survey and Euclid. Models are for the central cosmological parameter values fitting WMAP+BAO+SNe described in \S1, using the $\Lambda$CDM background (for $\Lambda$CDM and $f(R)$) and the DGP background (for DGP).}
 \label{fig:correlation lcdm}
\end{figure*}

\begin{figure*}
\centering
\mbox{
\subfigure[Including non-linear effects for sources with $z_{\rm m}=0.825$ with ground-based
errors]{\psfig{figure=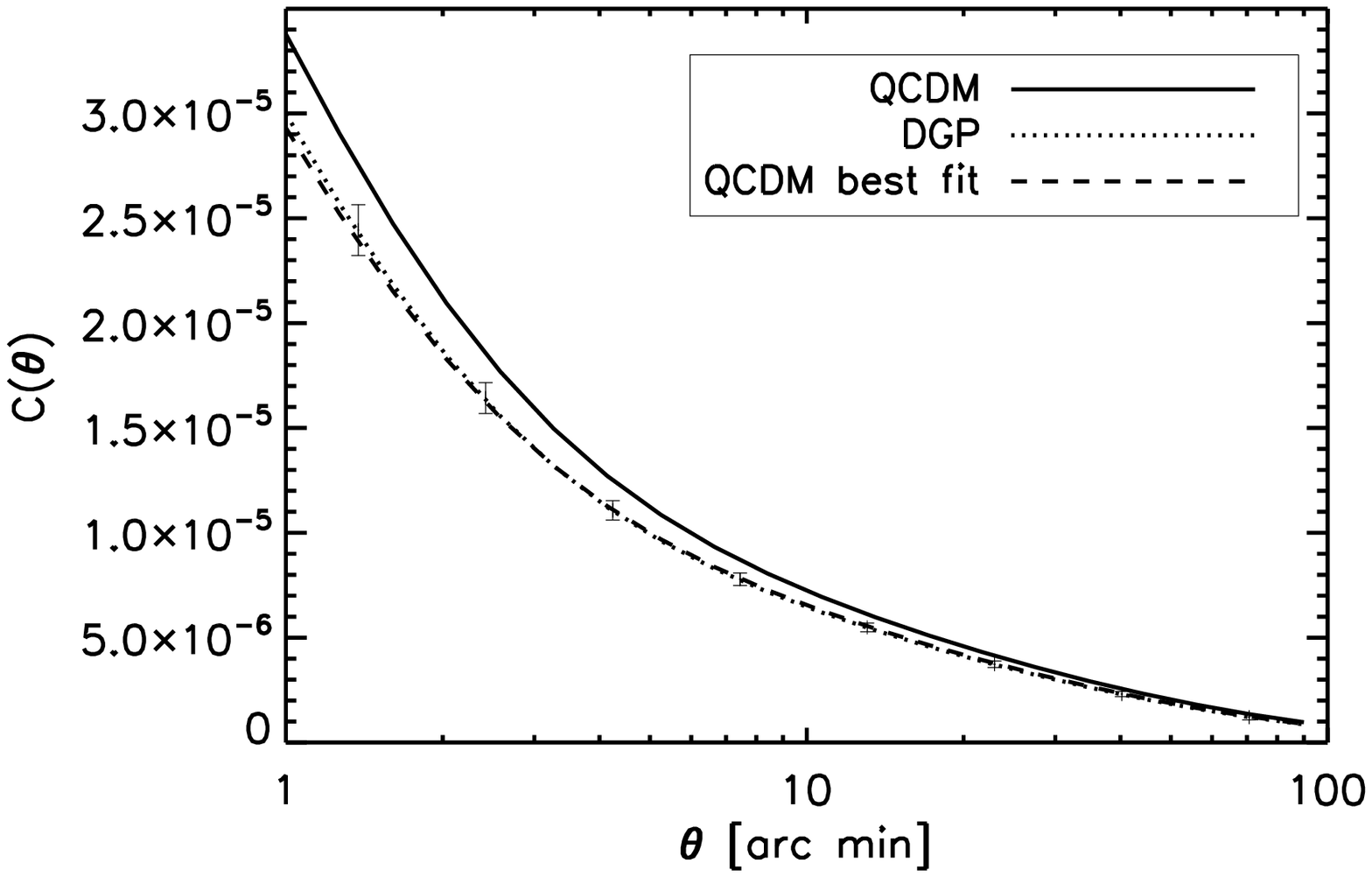,width=84mm}}
}
\mbox{
\subfigure[Including non-linear effects for sources with $z_{\rm m}=0.9$ with Euclid
errors]{\psfig{figure=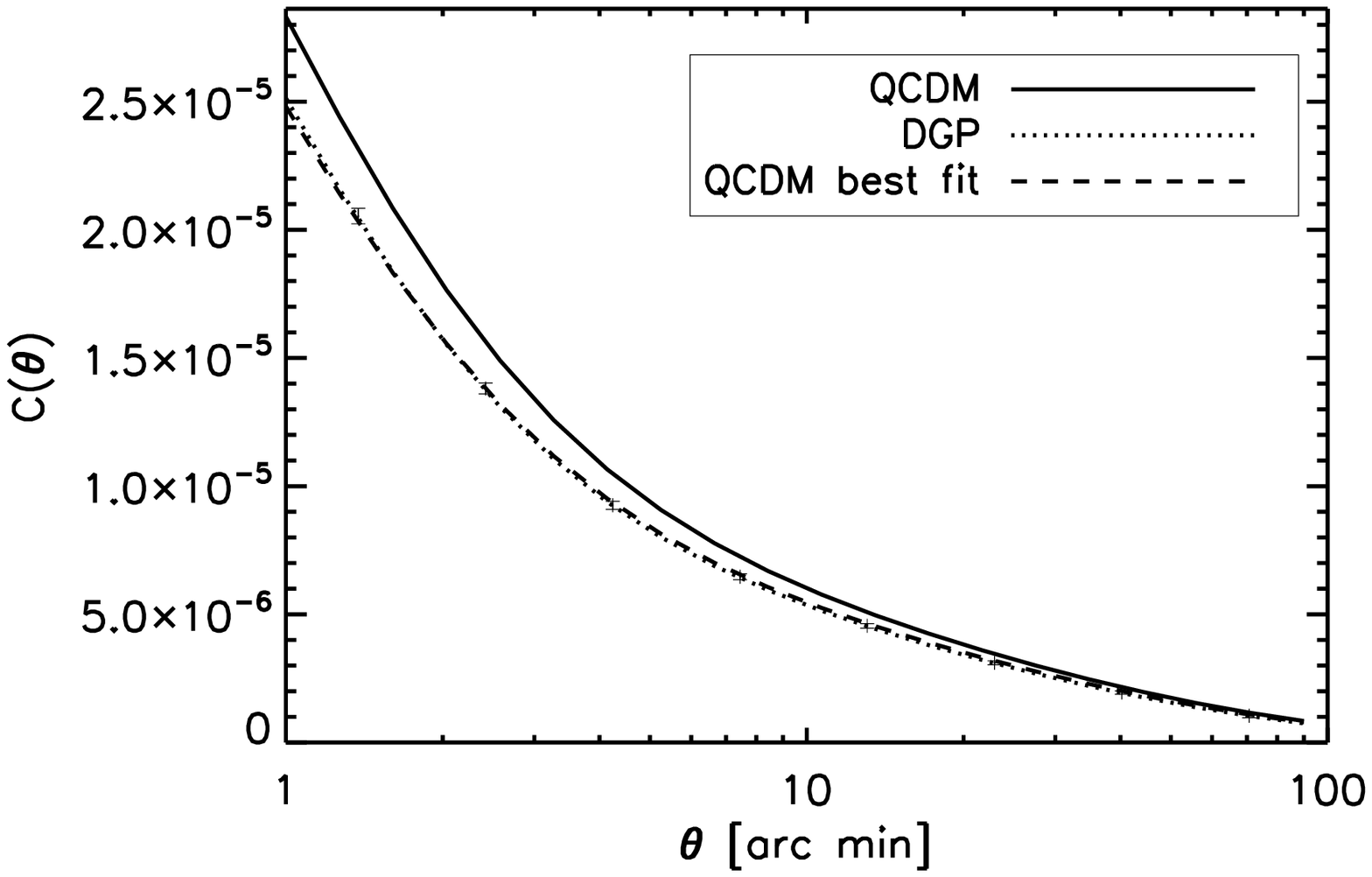,width=84mm}}
}
 \caption{Correlation function predicted for the QCDM model with the expansion history as DGP and DGP with error estimates for ground-based survey and Euclid. The solid lines show the correlation function for the QCDM model for the central cosmological parameter values fitting WMAP+BAO+SNe, using the DGP background. The dashed line shows the best fit QCDM model to the DGP model obtained by varying $\Omega_{\rm m}$ and $\sigma_8$.}
 \label{fig:correlation qcdm}
\end{figure*}

\begin{figure*}
\centering
\mbox{
\subfigure[$\Lambda CDM$ for $z=0.3-0.7$ redshift bin]{\psfig{figure=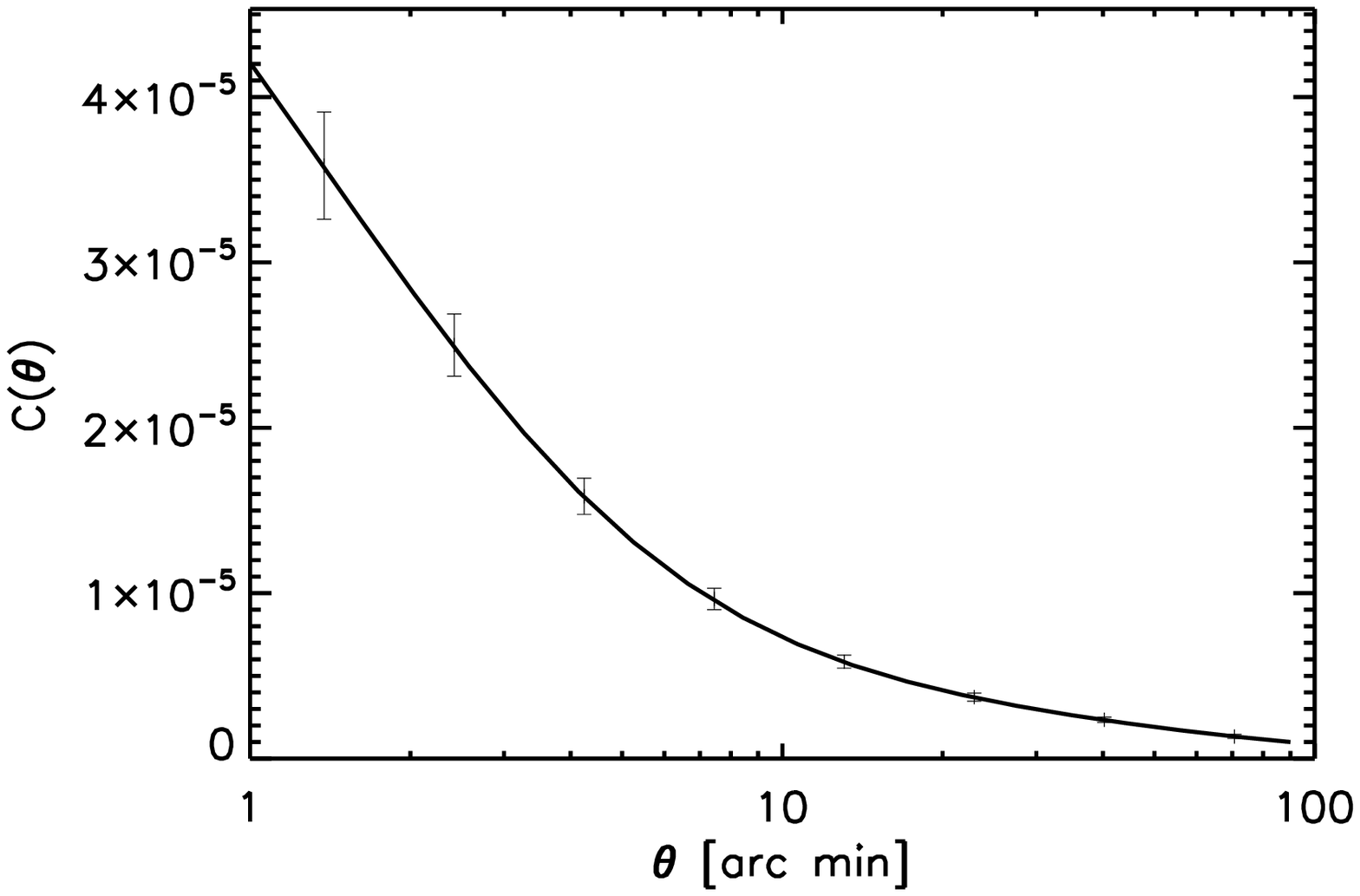,width=55mm}} \quad
\subfigure[DGP for $z=0.3-0.7$ redshift bin]{\psfig{figure=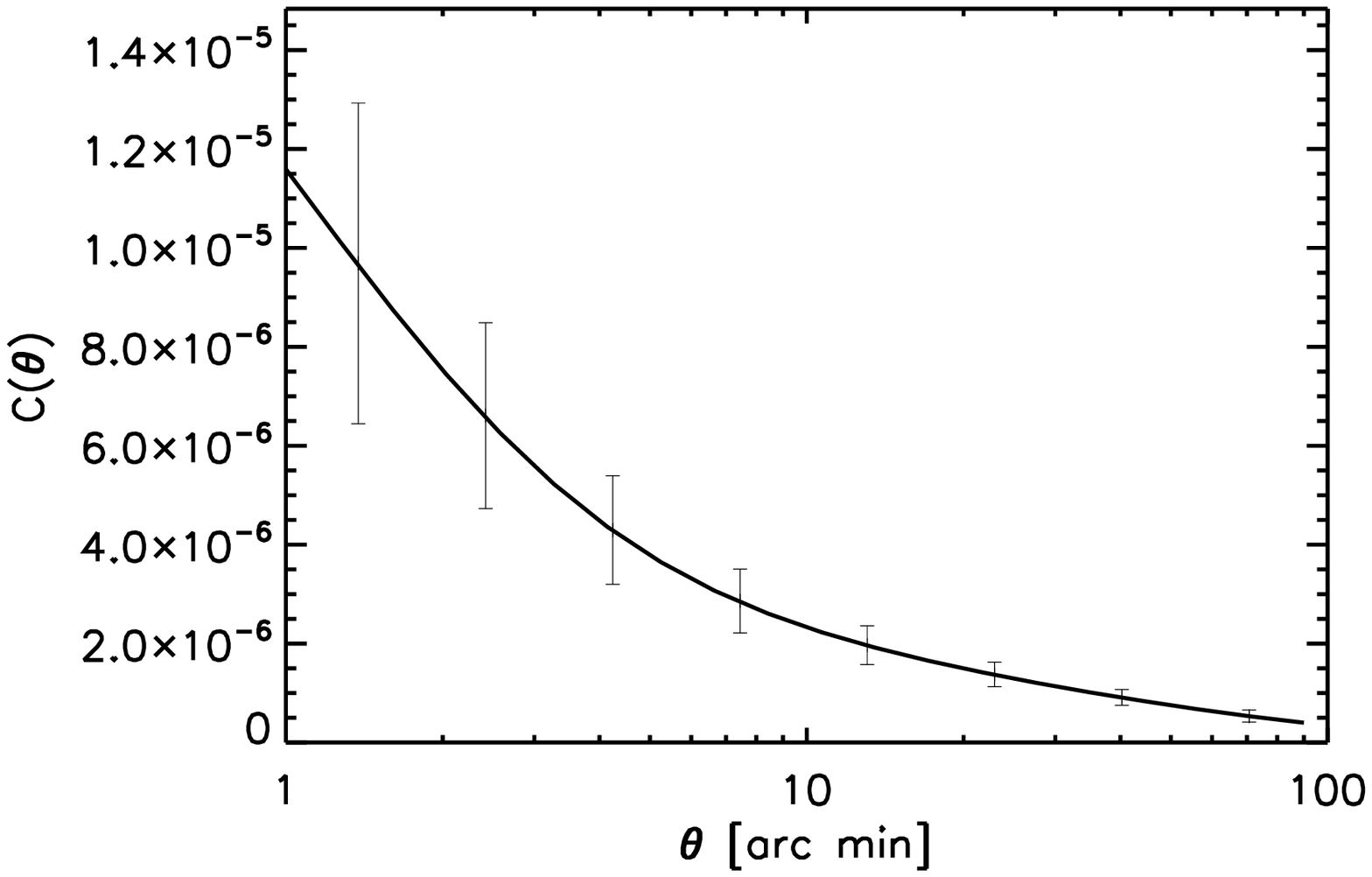,width=55mm}} \quad
\subfigure[$f(R)$ for $z=0.3-0.7$ redshift bin]{\psfig{figure=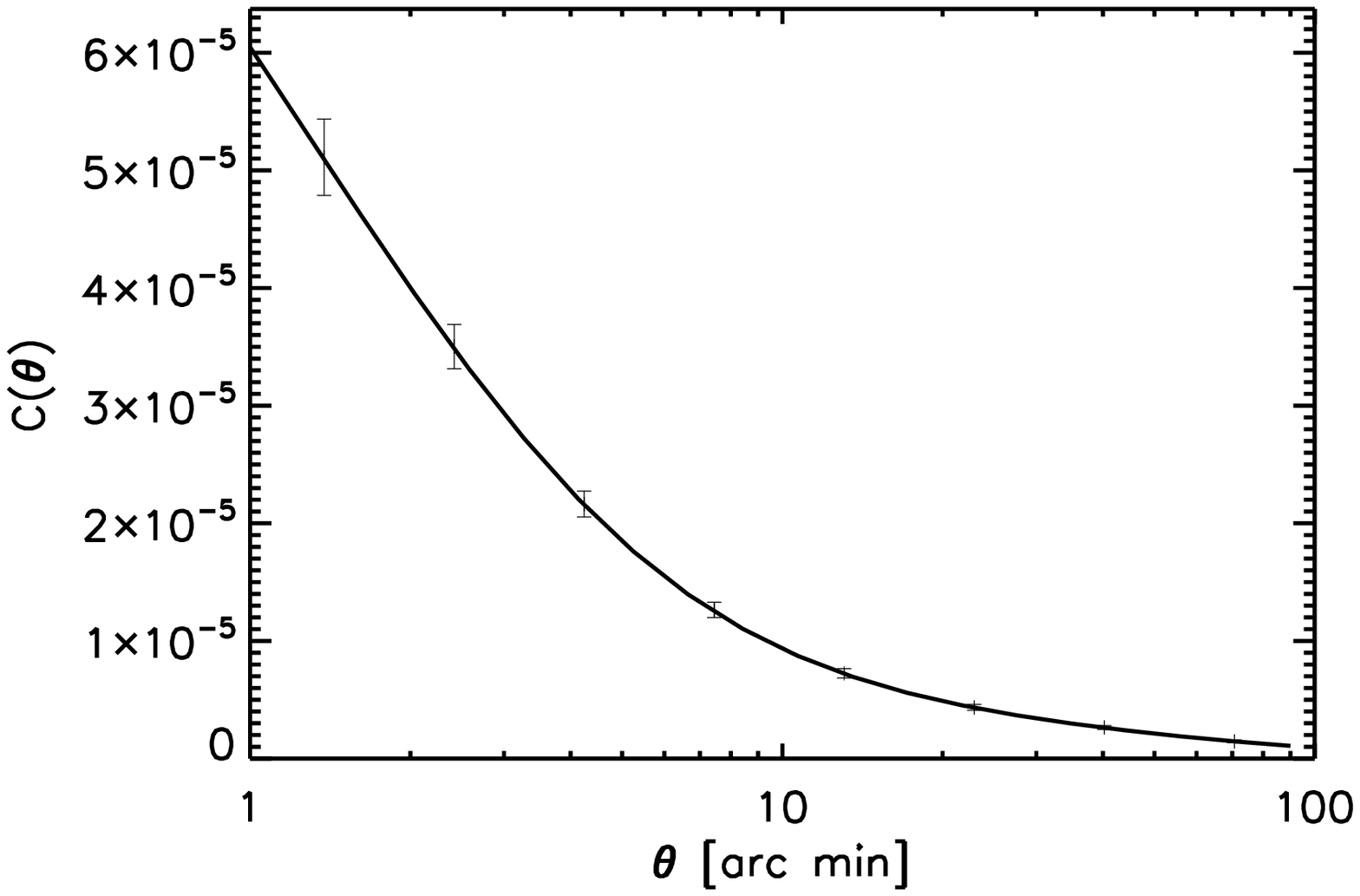,width=55mm}}
}
\mbox{
\subfigure[$\Lambda CDM$ for $z=0.7-1.1$ redshift bin]{\psfig{figure=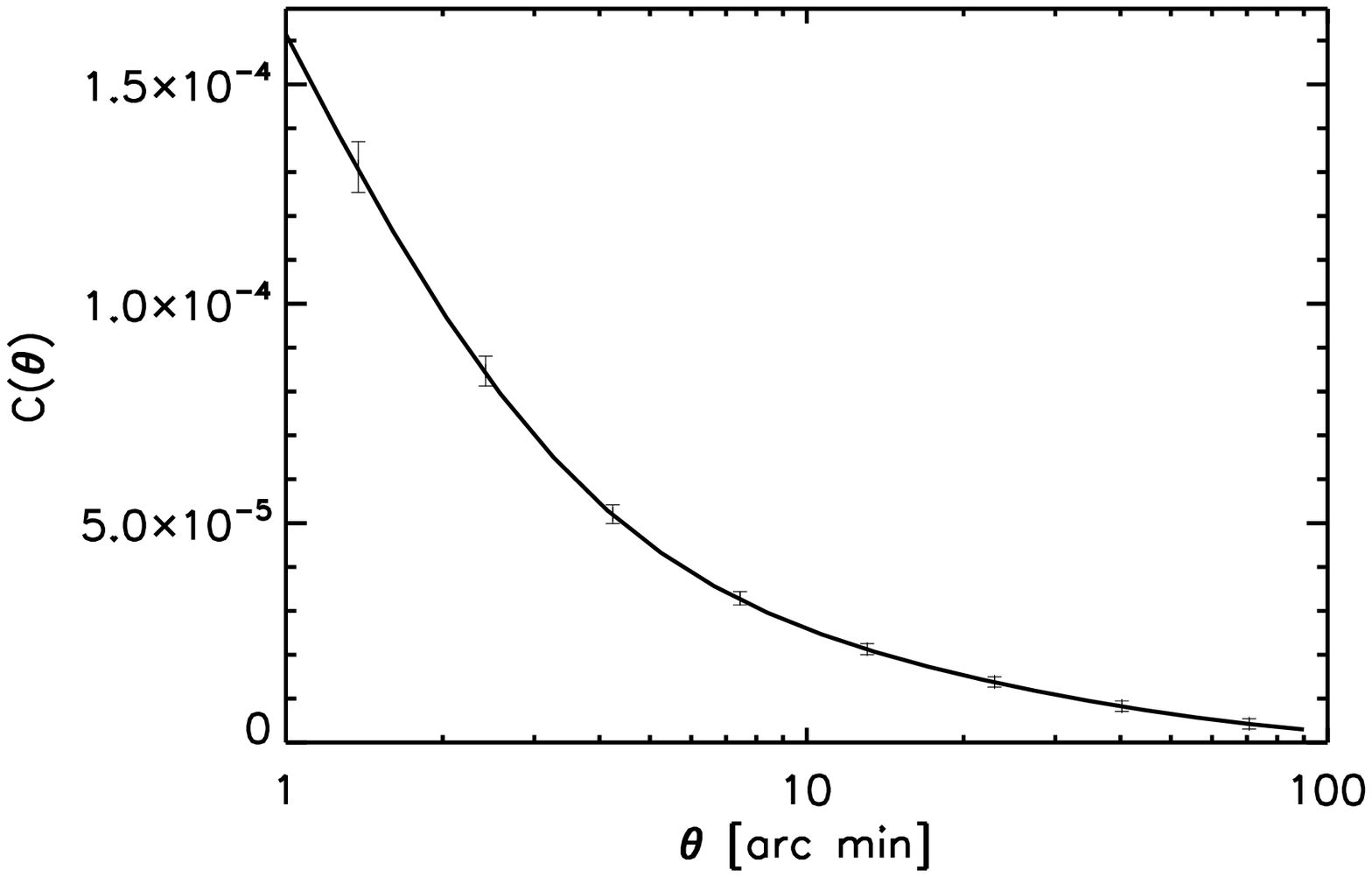,width=55mm}} \quad
\subfigure[DGP for $z=0.7-1.1$ redshift bin]{\psfig{figure=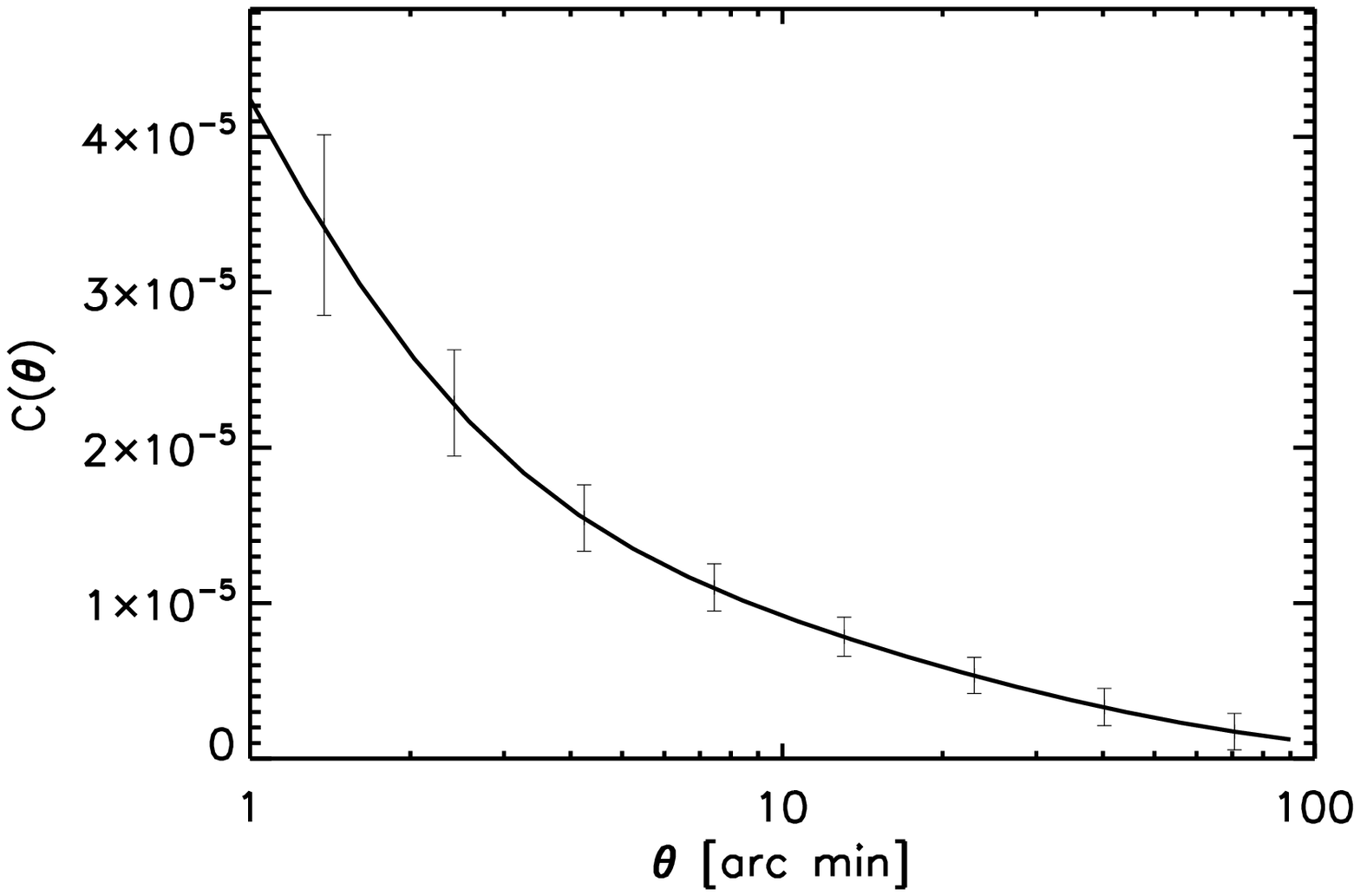,width=55mm}} \quad
\subfigure[$f(R)$ for $z=0.7-1.1$ redshift bin]{\psfig{figure=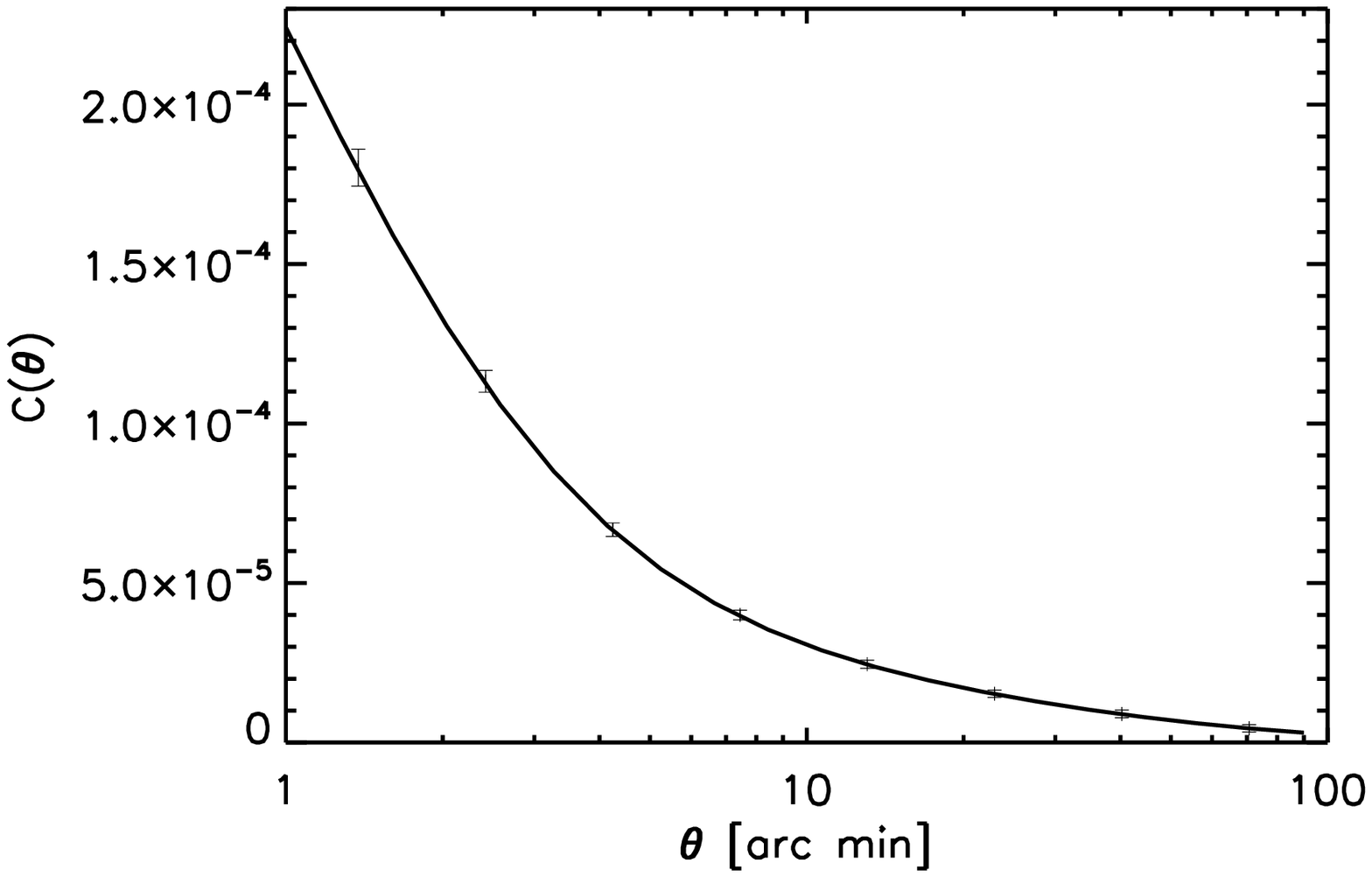,width=55mm}}
}
 \caption{Correlation function predicted for $\Lambda$CDM, DGP and $f(R)$ with error estimates for ground-based survey at different $z$ using redshift bins with width $\Delta z = 0.4$.}
 \label{fig:correlation different z}
\end{figure*}

Figure \ref{fig:correlation lcdm} shows example results for our ground-based survey and Euclid using the central cosmological parameter values for WMAP+SNe+BAO described in \S1; this is for the 2-D projection case where we have not divided the catalogue tomographically. We see from figures (a) and (b) that the difference between models is substantially greater in the nonlinear regime ($\theta \la 30'$) than in the linear regime ($\theta \ga 30'$), as is the amplitude of the signal. As (c) and (d) show, it is also the case that the linear correlation function is small in the low-$\theta$ regime, if nonlinear corrections are not included.

We present the $\chi^2$ differences between the modified gravities and fiducial dark energy models in Table \ref{chi sq table}, for the 2-D (non-tomographic) cases including non-linear power. We see that there is indeed strong discriminatory power between modified gravity models and $\Lambda$CDM with the notional ground-based survey; the precision of Euclid is even more impressive.

We also compare the constraints on DGP and a QCDM model of the same expansion history (i.e. a DGP background). The correlation functions for these models are shown in Figure \ref{fig:correlation qcdm}. One can either consider a QCDM model with cosmological parameters equal to their central values in a fit to WMAP+BAO+SNe, or more realistically the best fit QCDM model to the DGP model obtained by varying $\Omega_{\rm m}$ and $\sigma_8$.
We see that there is a choice of $\Omega_m$ and $\sigma_8$ that make the QCDM and DGP models virtually indistinguishable. This is confirmed by the bottom row of Table \ref{chi sq table}, which shows that the difference in $\chi^2$ for DGP and this QCDM is insignificant. This is  clearly partly due to the existence of a QCDM model with rather similar growth to the DGP, but also because of the low amplitude of the DGP correlation function, with the result that the error bars are larger in proportion to the signal than for other models. 

The power of future surveys to discriminate between gravity models is borne out by the tomographic results. Examples of these are shown in Figure \ref{fig:correlation different z}, where we see the different redshift evolutions and amplitudes of the signal in the different gravities. Table \ref{chi sq table 2} confirms that using the redshift information affords us better discrimination between dark energy and modified gravity models in every case, by a factor of 50 to 100\%. Because of this, we will only consider tomographic results from now on in the paper.

\begin{table}
\centering
\begin{tabular}{cccc}
	Fiducial & Modified & Ground-based & Euclid \\
	Model & gravity & $\Delta\chi^2$ & $\Delta\chi^2$\\
\hline
	\multirow{4}{*}{$\Lambda$CDM} & DGP & $4\times 10^3$ & $3\times 10^4$\\
	& $f(R)$, $\alpha_1=1/3$ & 500 & $6\times 10^3$\\
	& $f(R)$, $\alpha_1=1$ & 200 & $2\times 10^3$ \\
	& $f(R)$, $\alpha_1=2$ & 40 & 500\\
	\hline
  QCDM & DGP & 0.5 & 3\\
\end{tabular}
\caption{$\Delta\chi^2$ for DGP and $f(R)$ using errors from our ground-based survey and Euclid, with no redshift information, and using priors from WMAP$+$SNe$+$BAO. The top section shows results compared to $\Lambda$CDM, while the bottom row is compared to QCDM. }
\label{chi sq table}
\end{table}

\begin{table}
\centering
\begin{tabular}{cccc}
	Fiducial & Modified & Ground-based & Euclid \\
	Model & gravity & $\Delta\chi^2$ & $\Delta\chi^2$\\
	\hline
	\multirow{4}{*}{$\Lambda$CDM} & DGP & $6 \times 10^3$ & $7\times 10^4$ \\
	& $f(R)$, $\alpha_1=1/3$ & 600 & $8\times 10^3$ \\
	& $f(R)$, $\alpha_1=1$ & 300 & $3\times 10^3$ \\
	& $f(R)$, $\alpha_1=2$ & 60 & $1\times 10^3$\\
	\hline
  QCDM & DGP & 0.5 & 5 \\
\end{tabular}
\caption{Same as Table \ref{chi sq table}, but using tomographic information. In each case we have redshift bins of width $\Delta z = 0.4$ between $z=0.3$ and $1.5$.}
\label{chi sq table 2}
\end{table}

\begin{table}
\centering
\begin{tabular}{cccc}
	Fiducial & Modified & Ground-based & Euclid \\
	Model & gravity & $\Delta\chi^2$ & $\Delta\chi^2$\\
\hline
  \multirow{2}{*}{$\Lambda$CDM} & DGP & 500 & 3000 \\
	& $f(R)$ & 3 & 20 \\
	\hline
	QCDM & DGP & 0.2 & 2 \\
\end{tabular}
\caption{$\Delta\chi^2$ if only linear power is included for $\theta=30'-90'$, for 0.4 redshift bins between 0.3 and 1.5 using priors from WMAP$+$SNe$+$BAO.}
\label{linear non-linear table}
\end{table}

Table \ref{linear non-linear table} shows the impact of including non-linear power on our ability to discriminate between modified gravities. Comparing these results with Table \ref{chi sq table 2} we can see the improvement that measurements from the non-linear regime of the correlation functions provide. The improvement is very substantial, amounting to an order of magnitude in $\chi^2$ difference. 

\begin{table}
\centering
\begin{tabular}{cccc}
	Fiducial & Modified & Ground-based  & Euclid \\
  Model & gravity & \% difference & \% difference\\
	\hline
  \multirow{4}{*}{$\Lambda$CDM} & DGP & -3\% & -3\% \\
	& $f(R)$, $\alpha_1=1/3$ & -40\% & -40\% \\
	& $f(R)$, $\alpha_1=1$ & -70\% & -80\% \\
	& $f(R)$, $\alpha_1=2$ & -90\% & -90\% \\
	\hline
	QCDM & DGP & -70\% & -80\% \\
\end{tabular}
\caption{Percentage difference in $\Delta\chi^2$ if the Smith et al. (2003) formula is used with no attempt to fit GR at small scales, compared to using the Hu \& Sawicki fitting formula. All results are tomographic with WMAP+SNe+BAO priors as before.}
\label{smith fitting table}
\end{table}

It is important to note that using only the \cite{Smith:2002dz} formula, without the GR asymptote, causes an overestimation in our ability to discriminate between modified gravity and dark energy models as shown in Table \ref{smith fitting table}. This can amount to up to a 90\% difference in $\Delta \chi^2$ for some models, due to the difference in power at small scales that is present when there is no attempt to recover GR. This shows the importance of careful modelling of the nonlinear regime, including the appropriate small-scale GR limit.

\section{Parameterisation of the Power Spectrum}\label{parameterisation}

The sensitivity of lensing to changes in the matter power spectrum will be very important in determining the correct theory of gravity or dark energy in the near future. In this section we will therefore parameterise the non-linear power spectrum, in order to more fully understand what aspect of the power spectrum it is which lensing surveys will be sensitive to.

We use the growth factor $\gamma$ \citep{Linder:2005in} as is used in \citet{Amendola:2007rr}, but we also include the parameters used in the Hu and Sawicki fitting formula (Equation \ref{eq:PPF} and \ref{sigma}). In the formalism of \citet{Linder:2005in} the growth history, $g(a)$, is given by
\begin{equation}
	g(a)=\exp\left(\int^1_a \left[1-\left(\frac{\Omega_{\rm m}}{a^3H^2}\right)^\gamma \right]\frac{da}{a}\right),
\end{equation}
where $\gamma$ is set by the model. This parameterisation cannot model all theories of gravity, since it does not allow for growth histories which have $k$ dependency, such as $f(R)$. It is also only valid for gravity models where the combination of $\Phi + \Psi$ is the same as in GR, which is true for DGP \citep{Koyama:2006ef} and $f(R)$ for $f_{R0} \ll 1$ \citep{Oyaizu:2008tb}.

\begin{figure*}
\centering
\mbox{
\subfigure[$\Lambda$CDM fiducial model with ground-based errors]{\psfig{figure=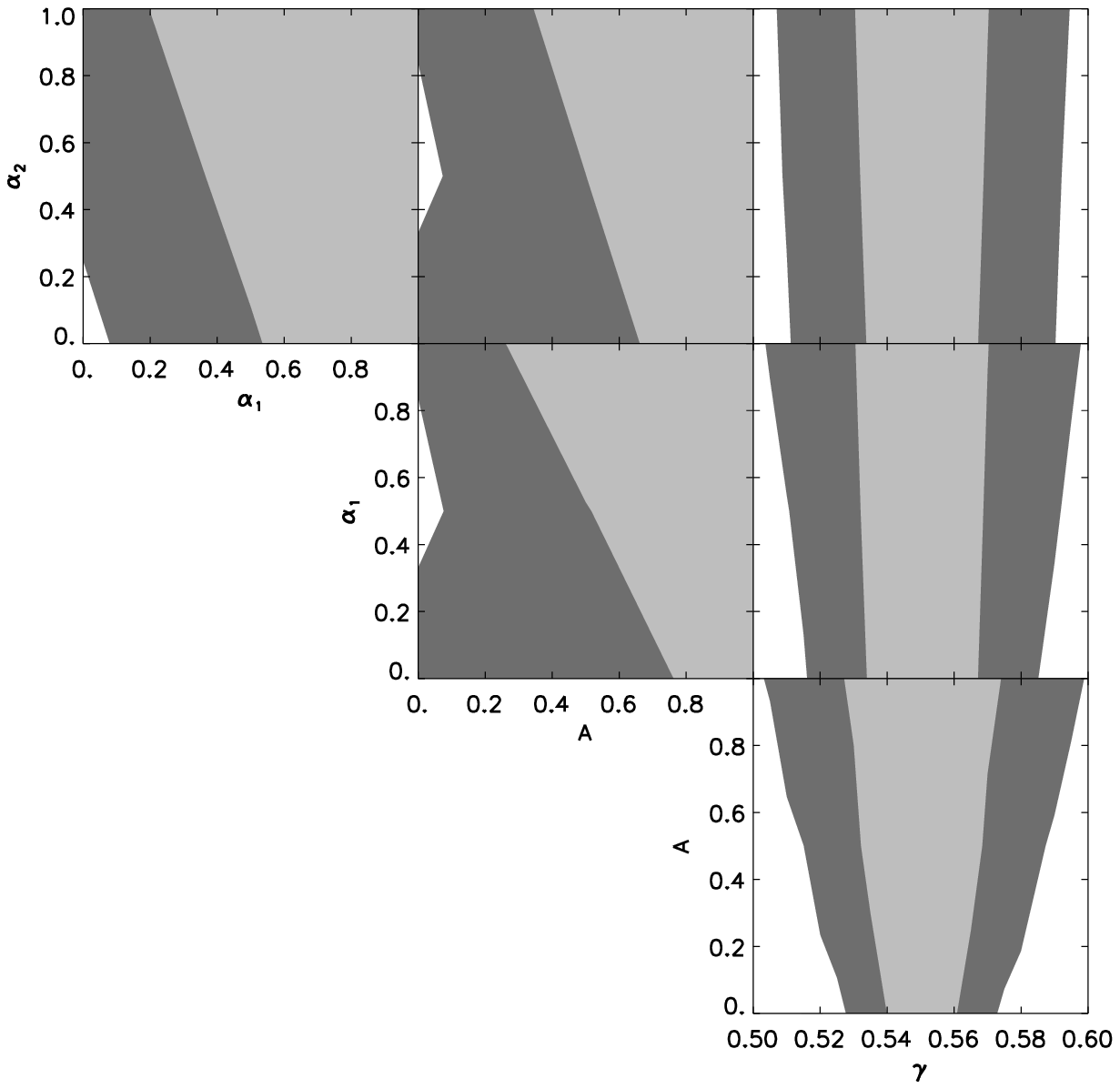,width=110mm} \label{fig:lcdm des central}} \hspace{-22 mm}
\subfigure[$\Lambda$CDM fiducial model with Euclid errors]{\psfig{figure=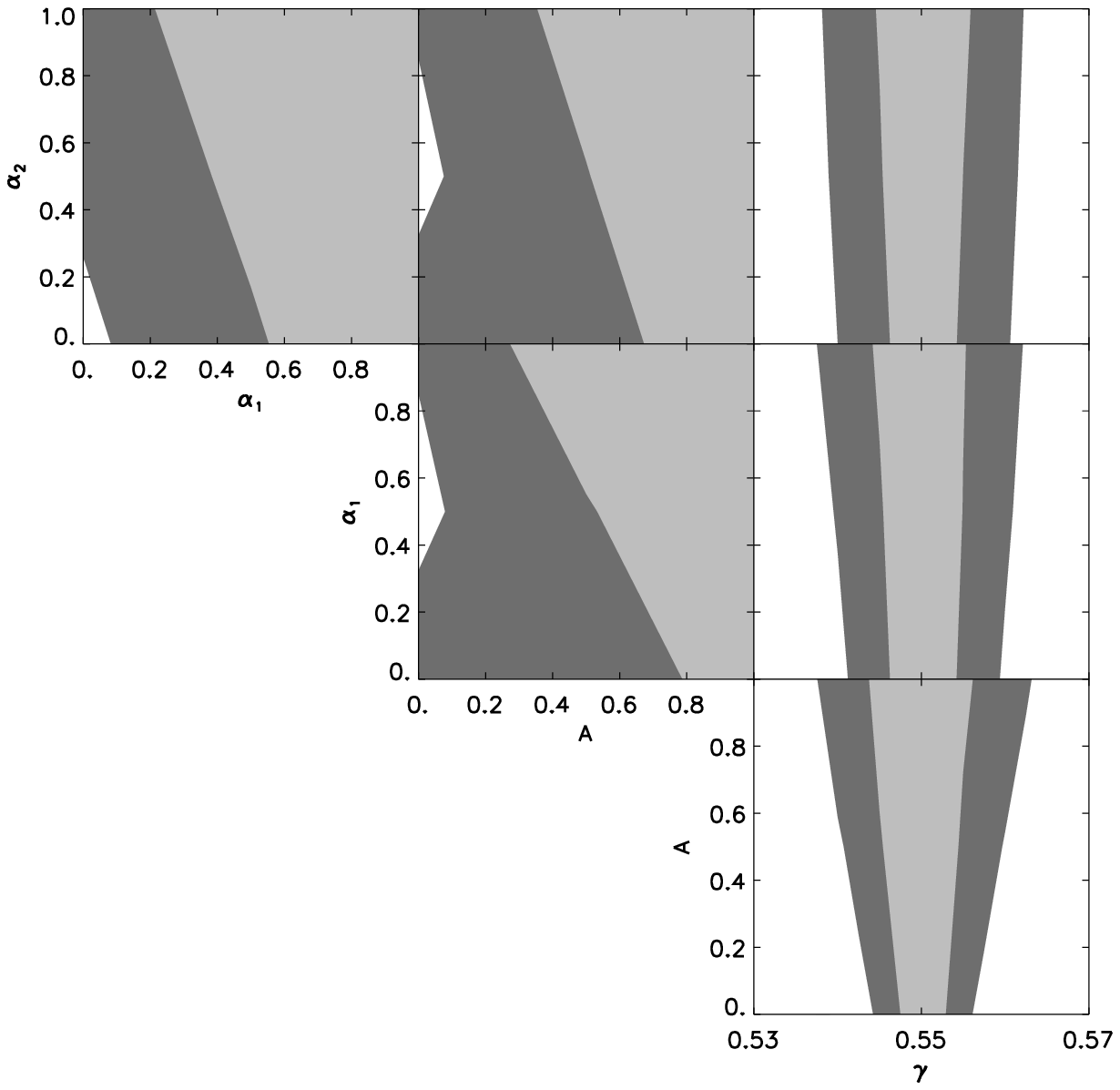,width=110mm} \label{fig:lcdm euclid central}}
}
\mbox{
\subfigure[DGP fiducial model with ground-based errors]{\psfig{figure=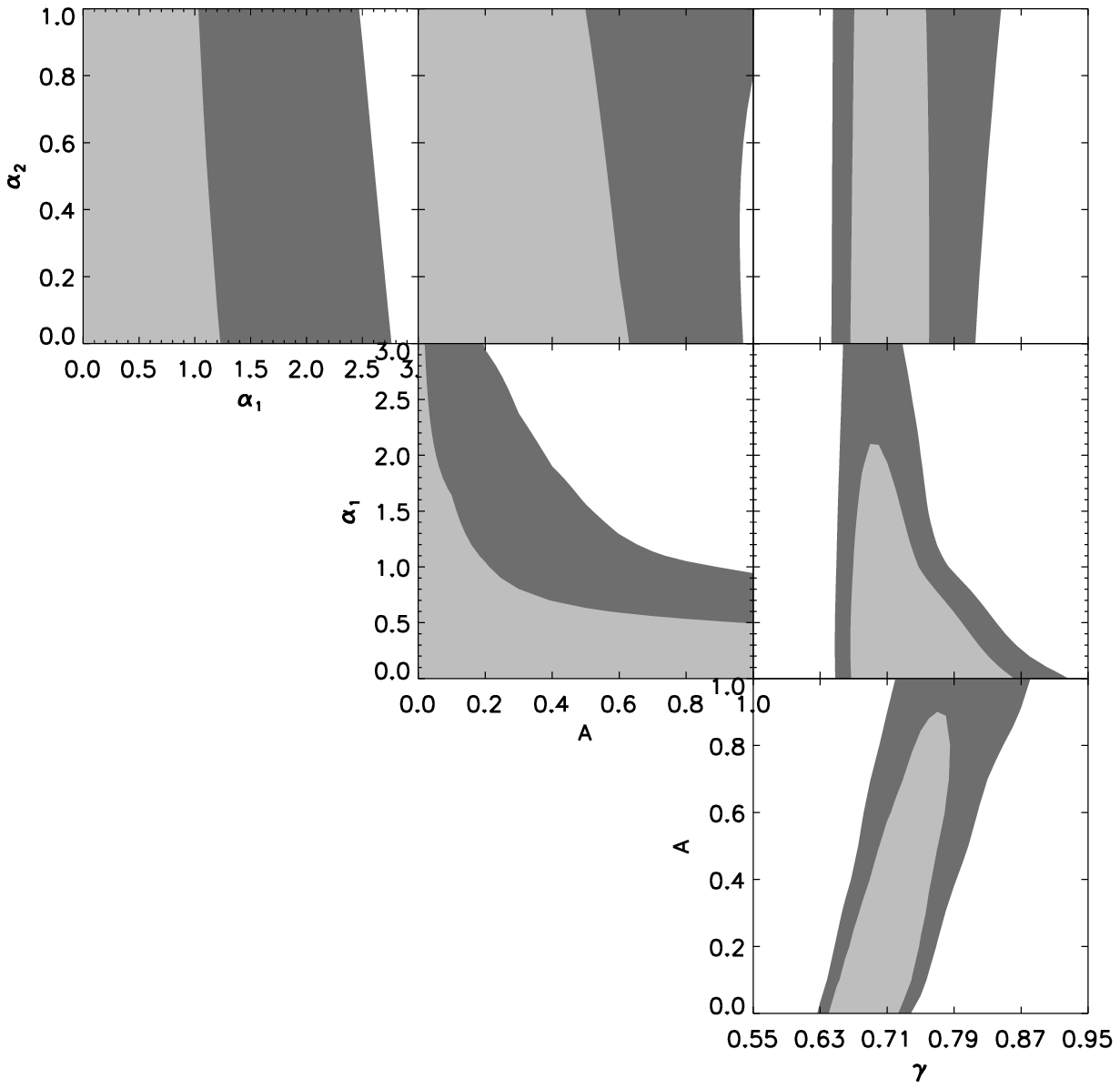,width=110mm} \label{fig:dgp des central}} \hspace{-22 mm}
\subfigure[DGP fiducial model with Euclid errors]{\psfig{figure=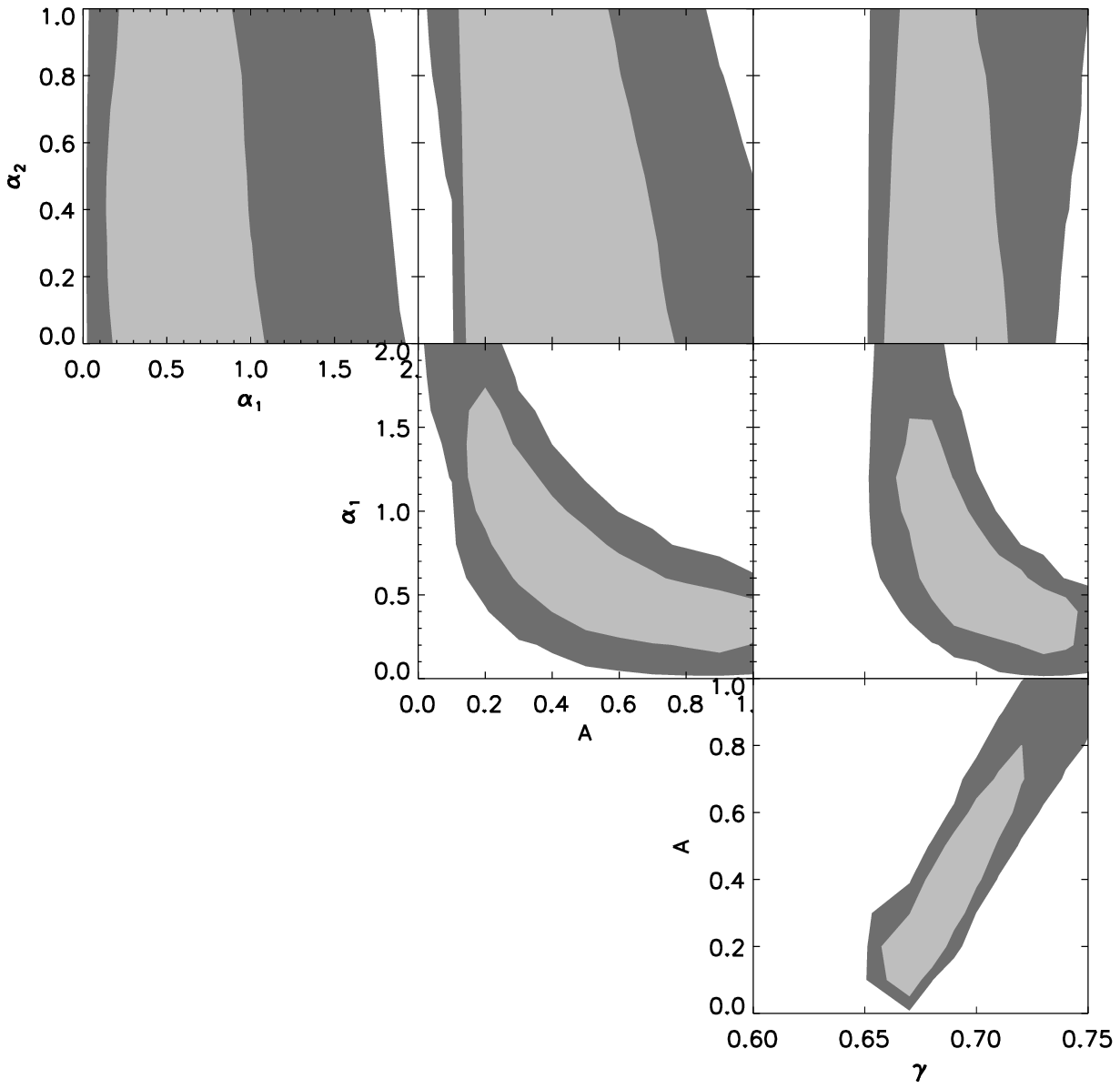,width=110mm} \label{fig:dgp euclid central}}
}

\caption{Constraints on $\gamma$, $\alpha_1$, $\alpha_2$ and $A$ from our ground-based survey and Euclid, using 0.4 redshift bins between 0.3 and 1.5 for the central cosmological parameter values fitting WMAP+BAO+SNe described in \S1. The light grey contours show the 68\% confidence limits and the dark grey show the 95\% confidence limits.}
\label{fig:Parameterisation single Om}
\end{figure*}

\begin{figure*}
\centering
\mbox{
\subfigure[$\Lambda$CDM fiducial model with ground-based errors]{\psfig{figure=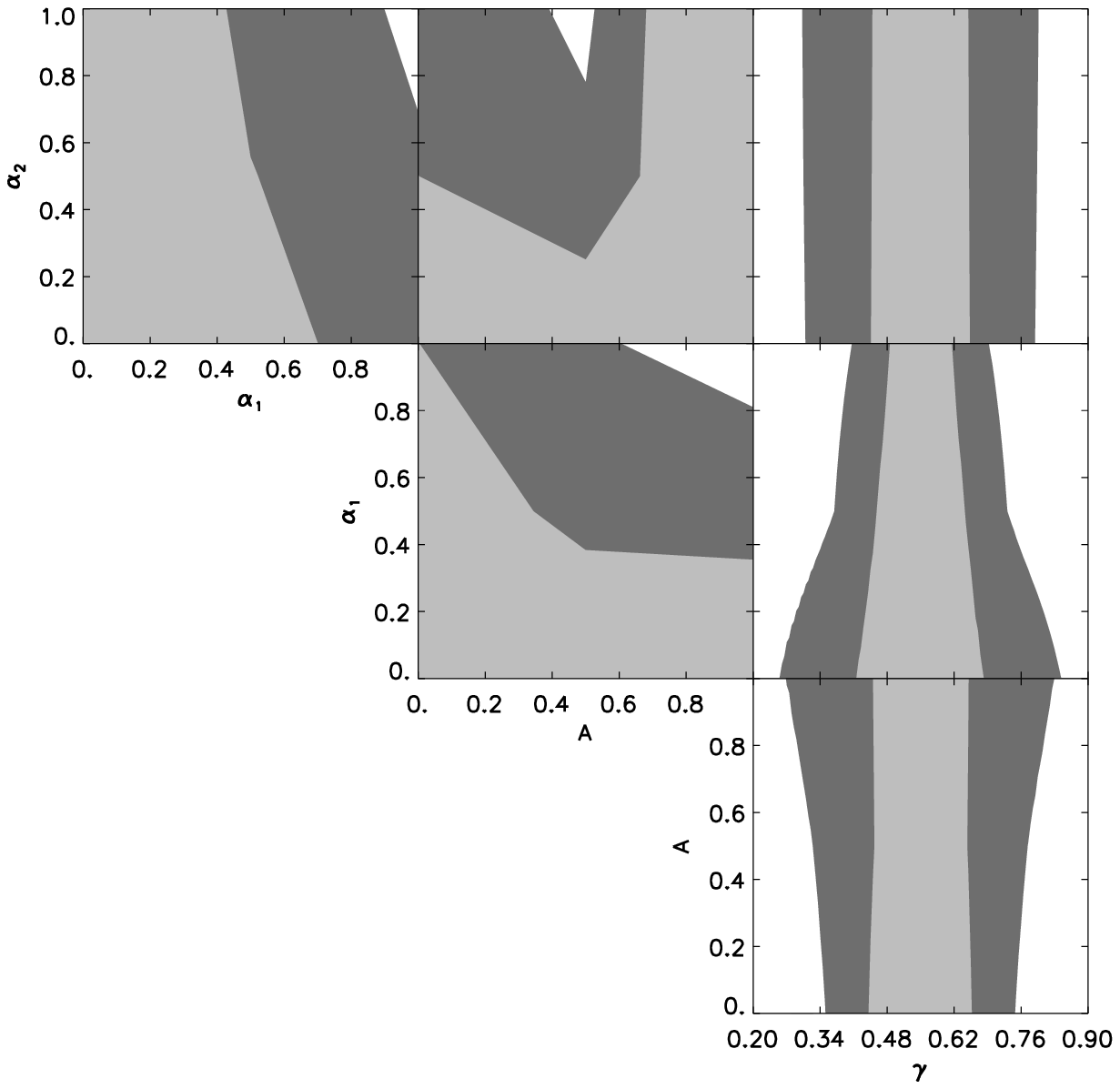,width=110mm} \label{fig:lcdm des param}} \hspace{-22 mm}
\subfigure[$\Lambda$CDM fiducial model with Euclid errors]{\psfig{figure=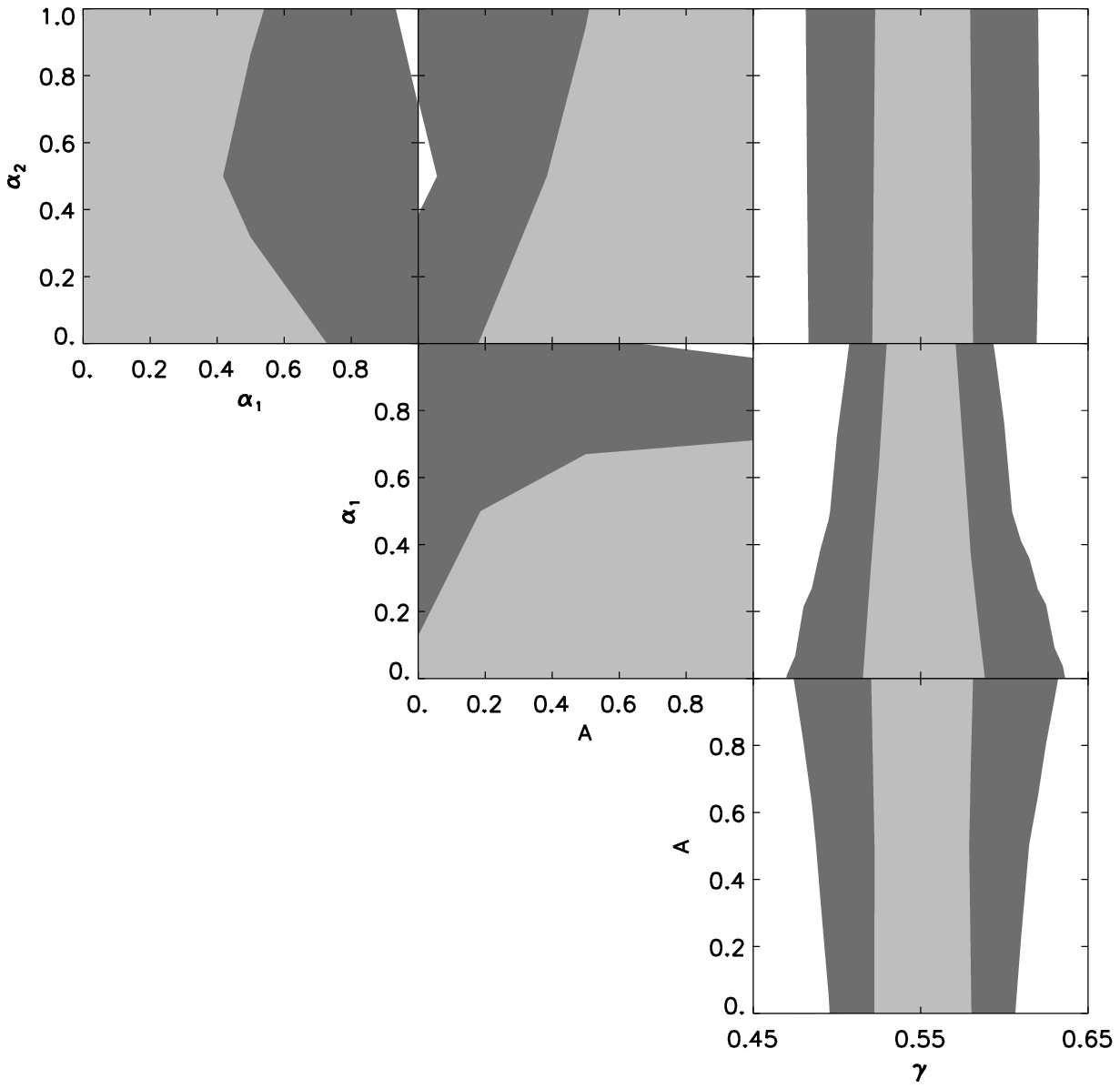,width=110mm} \label{fig:lcdm euclid param}}
}
\mbox{
\subfigure[DGP fiducial model with ground-based errors]{\psfig{figure=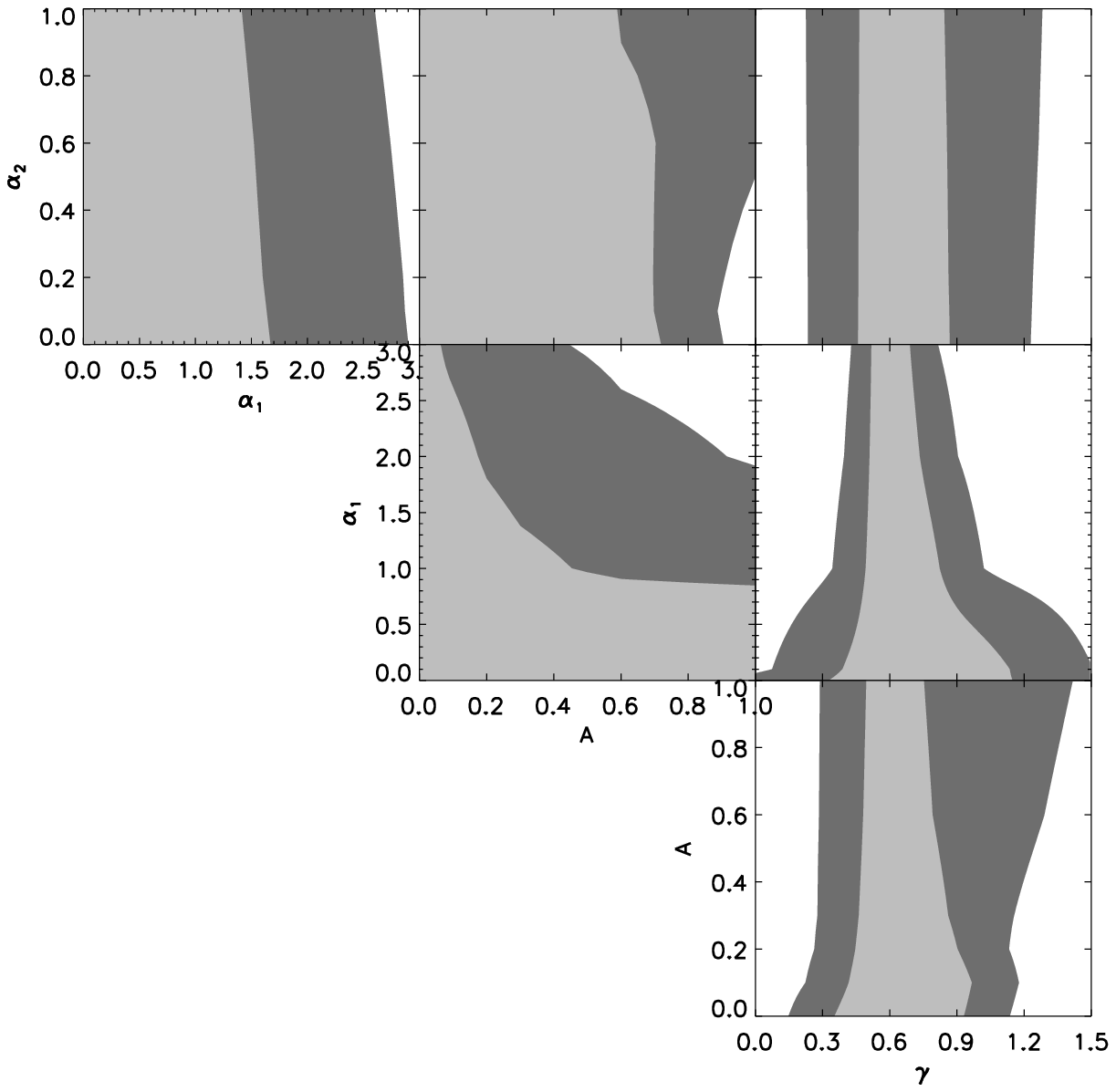,width=110mm} \label{fig:dgp des param}} \hspace{-22 mm}
\subfigure[DGP fiducial model with Euclid errors]{\psfig{figure=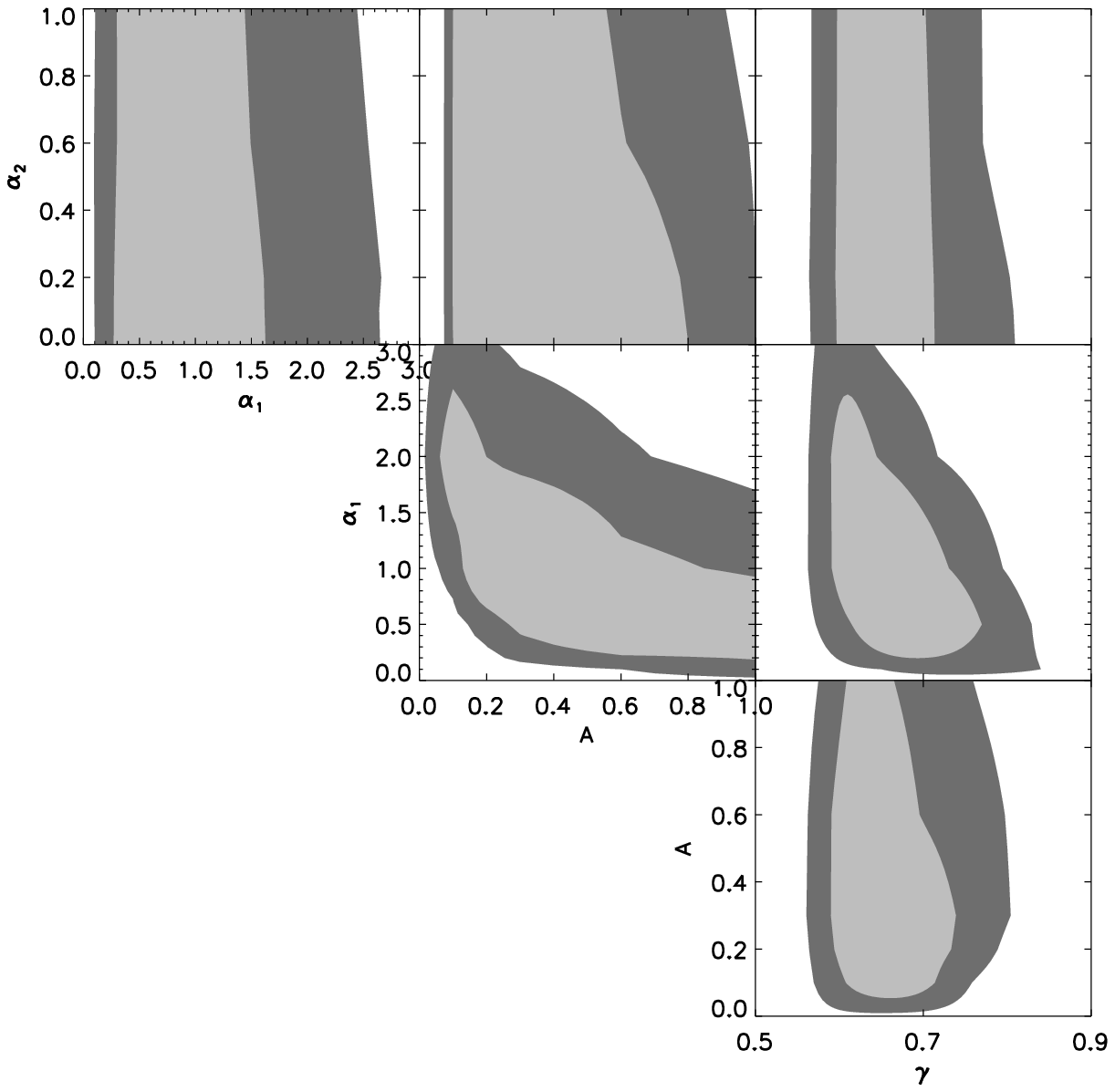,width=110mm} \label{fig:dgp euclid param}}
}

\caption{Constraints on $\gamma$, $\alpha_1$, $\alpha_2$ and $A$ from our ground-based survey and Euclid, using 0.4 redshift bins between 0.3 and 1.5, where we have marginalised over all $\Omega_m$ and $\sigma_8$. The light grey contours show the 68\% confidence limits and the dark grey show the 95\% confidence limits.}

\label{fig:Parameterisation}
\end{figure*}

Figures \ref{fig:lcdm des central} and \ref{fig:lcdm euclid central} demonstrate the dependence of the parameters on one another when fitting weak lensing predictions for varying $\gamma$, $A$, $\alpha_1$ and $\alpha_2$ to a $\Lambda$CDM fiducial model when $\Omega_m$ and $\sigma_8$ are fixed at the central values fitting WMAP+BAO+SNe. The slight widening in the $\gamma$ constraint as $A$, $\alpha_1$ and $\alpha_2$ increase is due to being able to recover $\Lambda$CDM at non-linear scales by increasing $A$ and $\alpha_1$ as $\gamma$ varies. This means that the constraint on $\gamma$ degrades slightly by including the parameters in the Hu and Sawicki fitting formula ($A$, $\alpha_1$ and $\alpha_2$). The constraint obtained by marginalising over all $\Omega_m$ and $\sigma_8$ shown in Figures \ref{fig:lcdm des param} and \ref{fig:lcdm euclid param} shows that the constraint for $\gamma$ for a $\Lambda$CDM fiducial model is very good, as shown in Table \ref{sigma1 sigma2 lcdm}, measuring $\gamma$ within 20\% of its value for the ground-based survey and within 5\% for Euclid, while the other parameters are difficult to constrain.

A better constraint on the parameters can be found for a growth history that is not $\Lambda$CDM, such as DGP, as shown in Figures \ref{fig:dgp des param} and \ref{fig:dgp euclid param}. This provides a better constraint on $A$, $\alpha_1$ and $\alpha_2$, but the constraint on $\gamma$ is not as tight, as shown in Table \ref{sigma1 sigma2 dgp}, measuring $\gamma$ within 30\% of its value for the ground-based survey and within 12\% for Euclid. This is due to the degeneracy between $\gamma$ and the other parameters in this instance. These degeneracies can be seen more clearly before the results are marginalised over $\Omega_m$ and $\sigma_8$ as shown in Figures \ref{fig:dgp des central} and \ref{fig:dgp euclid central}. The large dependence on the other fitting parameters demonstrates that care should be taken when predicting $\gamma$ constraints using this parameterisation.

\begin{table}
\centering
\begin{tabular}{ccccc}
	 &  &  Our & & Smith et al. \\
	Survey &  & parameterisation & Linear & (2003) \\
	\hline
  Ground & 68\% & 0.10 & 0.23 & 0.091 \\
	-based & 95\% & 0.24 & 0.42 & 0.18 \\
	\multirow{2}{*}{Euclid} & 68\% & 0.030 & 0.12 & 0.026 \\
	& 95\% & 0.069 & 0.23 & 0.051 \\
\end{tabular}
\caption{The 68\% and 95\% confidence limits for the growth factor $\gamma$ obtained for our parameterisation with $\Lambda$CDM as the fiducial model compared to those obtained using only linear scales and compared to the constraint from using Smith et al. 2003 to model the non-linear. These are marginalised over $\Omega_m$, $\sigma_8$, $A$, $\alpha_1$ and $\alpha_2$.}
\label{sigma1 sigma2 lcdm}
\end{table}

\begin{table}
\centering
\begin{tabular}{ccccc}
	 &  &  Our & & Smith et al. \\
	Survey &  & parameterisation & Linear & (2003) \\
	\hline
  Ground & 68\% & 0.22 & 0.38  & 0.25 \\
	-based & 95\% & 0.59 & 0.68 &  0.48 \\
	\multirow{2}{*}{Euclid} & 68\% & 0.082 & 0.20 & 0.052 \\
	& 95\% & 0.12 & 0.39 & 0.10 \\
\end{tabular}
\caption{The 68\% and 95\% confidence limits for the growth factor $\gamma$ obtained for our parameterisation with DGP as the fiducial model compared to those obtained using only linear scales and compared to the constraint from using Smith et al. 2003 to model the non-linear. These are marginalised over $\Omega_m$, $\sigma_8$, $A$, $\alpha_1$ and $\alpha_2$.}
\label{sigma1 sigma2 dgp}
\end{table}

One might think then that it is better not to include non-linear scales and constrain only $\gamma$ on linear scales. However, there is substantial extra signal coming from the non-linear regime. In fact with our parameterisations, Tables \ref{sigma1 sigma2 lcdm} and \ref{sigma1 sigma2 dgp} show the percentage difference between the 68\% constraint obtained for $\gamma$ if only a linear analysis is used compared to the full non-linear analysis with the fitting formula is 100\% for the ground-based survey and 300\% for Euclid with a $\Lambda$CDM fiducial model and  70\% for the ground-based survey and 140\% for Euclid with a DGP fiducial model. 

The percentage overestimation, shown in Tables \ref{sigma1 sigma2 lcdm} and \ref{sigma1 sigma2 dgp}, at the 68\% level, in the ability of the ground-based survey and Euclid to constrain $\gamma$ if only the Smith et. al. fitting formula is used is 10\% for the ground-based survey and 40\% for Euclid with a $\Lambda$CDM fiducial model and 10\% for ground-based survey and 60\% for Euclid with a DGP fiducial model. This demonstrates that if a full non-linear analysis is to be used then it is necessary to ensure that GR is obtained at small scales, and the extra parameters from the Hu and Sawicki fitting formula must also be measured.

\section{Conclusions}\label{conclusions}

In this paper we have presented weak lensing predictions for modified gravity models, including the non-linear regime of the power spectrum.

We have shown how the power spectrum is calculated for DGP, $f(R)$ and QCDM models, using the fitting function of \cite{Hu:2007pj} to explore deep into the non-linear regime, while including the fact that gravities should tend towards GR on small scales.

We have calculated the total shear power spectrum given the modified gravity power spectrum, and have shown that this will be measured with high signal-to-noise with future lensing surveys such as Euclid and DES. We have taken into account the cosmic covariance in addition to the noise due to the intrinsic shapes of galaxies. 

We have shown that there is substantial additional discriminatory power between modified gravity models which is now afforded to us by the inclusion of the nonlinear power regime. We have also shown that using only the \cite{Smith:2002dz} formula without any attempt to obtain the GR non-linear power spectrum on small scales leads to an overestimatation in the ability of future surveys to differentiate between different growth histories.

We have parameterised the dark matter power spectrum using the growth factor $\gamma$ and the parameters in the non-linear fitting function to see how well a ground-based survey similar to DES, and a space-based survey such as Euclid, will be able to put constraints on these. We have compared the results from this parameterisation with results obtained from using only linear scales and have shown the constraint on $\gamma$ to be much tighter in the former case.

\section*{Acknowledgements}
We are very grateful to the Horizon project \citep{Teyssier:2009zd} for allowing us to use their simulation data to calculate our errors due to cosmic variance. We thank Takashi Hiramatsu for confirming the $\Omega_{\rm m}$ and $\sigma_8$ dependence of $A$, $\alpha_1$ and $\alpha_2$ using perturbation theory. We would also like to thank Shaun Thomas, Alan Heavens, Sarah Bridle, Bhuvnesh Jain and Ben Hoyle for their comments. DB is supported by an STFC Advanced Fellowship and an RCUK Academic Fellowship. KK is supported by ERC, STFC and RCUK. EB is funded by an STFC PhD studentship.

\label{lastpage}

\end{document}